\documentclass{emulateapj}
\usepackage{amsmath}
\usepackage{amssymb}
\usepackage{amsthm}
\usepackage{array}
\usepackage{float}
\usepackage{graphicx}
\usepackage{subfigure}

\shorttitle{Multi-Messenger Tests for Fast-Spinning Newborn NSs in SE SNe}
\shortauthors{Kashiyama et al.}

\begin{document}

\title{Multi-Messenger Tests for Fast-Spinning Newborn Pulsars \\ Embedded in Stripped-Envelope Supernovae}

\author{Kazumi Kashiyama$^{1}$, Kohta Murase$^{2,3}$, Imre Bartos$^{4}$, Kenta Kiuchi$^{5}$, and Raffaella Margutti$^{6}$}

\altaffiltext{1}{Einstein fellow---Department of Astronomy, Department of Physics, Theoretical Astrophysics Center, University of California, Berkeley, Berkeley, CA 94720, USA}
\altaffiltext{2}{School of Natural Sciences, Institute for Advanced Study, 1 Einstein Dr. Princeton NJ 08540}
\altaffiltext{3}{Department of Physics, Department of Astronomy and Astrophysics, Center for Particle and Gravitational Astrophysics, Pennsylvania State University, University Park, PA 16802, USA}
\altaffiltext{4}{Department of Physics, Columbia Astrophysics Laboratory, Columbia University, New York, NY 10027, USA}
\altaffiltext{5}{Yukawa Institute for Theoretical Physics, Kyoto University, Kyoto, 606-8502, Japan}
\altaffiltext{6}{Harvard-Smithsonian Center for Astrophysics, 60 Garden St., Cambridge, MA 02138, USA}

\begin{abstract}
Fast-spinning strongly magnetized newborn neutron stars (NSs), including nascent magnetars, are popularly implemented as the engine of luminous stellar explosions. 
Here, we consider the scenario that they power various stripped-envelope (SE) supernovae (SNe), not only super-luminous SNe Ic but also broad-line (BL) SNe Ibc and possibly some ordinary supernovae Ibc. 
This scenario is also motivated by the hypothesis that Galactic magnetars largely originate from fast-spinning NSs as remnants of SE SNe.  
By consistently modeling the energy injection from magnetized wind and ${}^{56}$Ni decay, 
we show that proto-NSs with $\gtrsim 10 \ \rm ms$ rotation and poloidal magnetic field of $B_{\rm dip} \gtrsim 5 \times 10^{14} \ \rm G$ can be harbored in ordinary SNe Ibc. 
On the other hand, millisecond proto-NSs can solely power BL SNe Ibc if they are born with $B_{\rm dip} \gtrsim 5 \times 10^{14} \ \rm G$, 
and superluminous SNe Ic with $B_{\rm dip} \gtrsim 10^{13} \ \rm G$. 
Then, we study how multi-messenger emission can be used to discriminate such pulsar-driven SN models from other competitive scenarios.  
First, high-energy X-ray and gamma-ray emission from embryonic pulsar wind nebulae can probe the underlying newborn pulsar. 
Follow-up observations of SE SNe using {\it NuSTAR} $\sim 50-100 \rm \ days$ after the explosion is strongly encouraged for nearby objects. 
We also discuss possible effects of gravitational waves (GWs) on the spin-down of proto-NSs. 
If millisecond proto-NSs with $B_{\rm dip} < {\rm a \ few} \times 10^{13} \ \rm G$ emit GWs through, e.g., non-axisymmetric rotation deformed by the inner toroidal fields of $B_{\rm t} \gtrsim 10^{16} \ \rm G$, 
the GW signal can be detectable from ordinary SNe Ibc in the Virgo cluster by Advanced LIGO, Advanced Virgo, and KAGRA.
\end{abstract}


\section{Introduction}
Time-domain astronomy is rapidly expanding. 
Various transients are now being efficiently detected and newly discovered by survey facilities, 
{\it Swift}~\citep{Gehrels_et_al_2004}, {\it Fermi}~\citep{Atwood_et_al_2009}, the Palomar Transient Factory~\citep[PTF:][]{Law_et_al_2009} 
and the Panoramic Survey Telescope \& Rapid Response System~\citep[Pan-STARRS:][]{Hodapp_et_al_2004}. 
%wide-field-of-view satellites like {\it Swift}~\citep{Gehrels_et_al_2004} and {\it Fermi}~\citep{Atwood_et_al_2009}, ground-based telescopes 
%like the Palomar Tansient Factory~\citep[PTF:][]{Law_et_al_2009} and the Panoramic Survey Telescope \& Rapid Response System~\citep[Pan-STARRS:][]{Hodapp_et_al_2004}.  
Deeper follow-up observations are well functioning in radio to gamma-ray bands~\citep[e.g.,][]{Greiner_et_al_2008,Perley_et_al_2011,Abeysekara_et_al_2012}. 
Moreover, Cherenkov detectors like Super-Kamiokande~\citep{Ikeda_et_al_2007} and IceCube~\citep{Aartsen_et_al_2015} are now gazing at the neutrino sky,  
and the installation of gravitational-wave (GW) interferometers like Advanced LIGO~\citep{Harry_et_al_2010}, Advanced Virgo~\citep{Accadia_et_al_2011}, and KAGRA~\citep{Somiya_2012} will be completed soon.  
These new and upgraded facilities can potentially probe hidden engines of violent astrophysical phenomena.  

Some targets among the promising sources of GWs and neutrinos are fast-spinning strongly magnetized proto-neutron stars~(NSs) formed in collapsing stars. 
They have been proposed as the engines of luminous transients, e.g., long gamma-ray bursts~\citep[L-GRBs; ][]{Usov_1992,Thompson_1994,Blackman_Yi_1998,Zhang_Meszaros_2001,Thompson_et_al_2004,Metzger_et_al_2007,Buccianitini_et_al_2009} 
including their sub-class with low luminosity~\citep[LL-GRBs;][]{Mazzali_et_al_2006,Soderberg_et_al_2006,Toma_et_al_2007},  
broad-line type Ibc SNe~\citep[BL-SN Ibc:][]{Wheeler_et_al_2000,Thompson_et_al_2004,Woosley_2010}, 
hydrogen-poor superluminous SNe \citep[SL-SNe Ic;][]{Kasen_Bildsten_2010,Pastorello_et_al_2010,Quimby_et_al_2011,Inserra_et_al_2013,Nicholl_et_al_2013}~\citep[see also][]{Metzger_et_al_2015,Wang_et_al_2015}. 
The basic picture is that the rotational energy of proto-NS is extracted by the unipolar induction as magnetized wind or jet and is later dissipated by a physical process that has still to be constrained, 
resulting in luminous electromagnetic radiation. 
However, no observational finding has been able to conclusively validate the pulsar-driven scenario so far. %\footnote{
%These models are often called magnetar models. 
%An observational definition of magnetar is NSs characterized by non-thermal radiation exceeding the spin-down power. 
%Based on this definition, magnetar is not a right word to describe central engines spinning down by magnetic braking. 
%An alternative definition is that NSs with a stronger magnetic fields than that of the quantum-electrodynamics~(QED) limit, $B_{\rm QED} \sim 4.4 \times 10^{13} \ \rm G$.
%Even based on this definition, proto-NSs considered in pulsar-driven SN model for SL-SN Ic are not magnetars.}
The question is how to discriminate newborn pulsar engines for each type of transient by using ongoing and upcoming multi-messenger observations. 

In this paper, using a semi-analytical model shown in the Appendix, we consistently calculate the multi-messenger counterparts from fast-spinning strongly magnetized proto-NSs, 
focusing on the cases accompanied by stripped-envelope~(SE) SNe.
In the next section, we discuss another important motivation of our study: the possible connection between Galactic magnetars and pulsar-driven SE SNe. 
Then, we consider the SN counterpart and derive the parameter range of the pulsar-driven SN model consistent with the observed  SN Ibc, BL-SN Ibc and SLSN-Ic (Sec. 3).\footnote{
In this paper, we do not include SNe IIb, which are usually listed as part of the SE SN family.} 
In Sec. 4, we show the detectability of the multi-messenger counterparts including the pulsar wind nebular~(PWN) emission, GW emission, and neutrino emission. 
Based on the results, we discuss observational strategies for the multi-messenger search of fast-spinning newborn NSs in SE SNe and possible scientific impacts in Sec. 5. 
We summarize our paper in Sec. 6. 

\section{Connection between Galactic Magnetars and Stripped-Envelope Supernovae?}
Confirming pulsar-driven scenarios is important in terms of understanding the origin of Galactic magnetars. 
In the classical picture~\citep{Duncan_Thompson_1992,Thompson_Duncan_1993}, the magnetic field amplification is attributed to the proto-NS convection coupled with a differential rotation less than $\rm a \ few \ ms$. 
Even in the absence of such rapid rotation, the magnetic fields could be amplified by the magnetorotational instability~\citep[e.g.,][]{Balbus_Hawley_1998,Akiyama_et_al_2003,Thompson_et_al_2005,Mosta_et_al_2015}. 
The total magnetic field energy of a magnetar is estimated to be:
\begin{equation}
{\cal E}_{\rm B} \approx B_{\rm t}{}^2/ 8\pi \times 4\pi R_{\rm ns}{}^3/3 \sim 2.9 \times 10^{49} \ {\rm erg} \ \left(\frac{B_{\rm t}}{10^{16} \ \rm G}\right)^2, 
\end{equation}
while the free rotational energy stored in the proto-NS is:
\begin{equation}\label{eq:E_rot}
{\cal E}_{\rm rot, i} \approx  I (2 \pi/P_{\rm i})^2/2 \sim 3.1 \times 10^{49} \ \left(\frac{P_{\rm i}}{30 \ \rm ms}\right)^{-2} \ \rm erg. 
\end{equation}
Here, $B_{\rm t}$ is the inner toroidal field strength, $I \sim 1.4 \times 10^{45} \ \rm g \ cm^2$ is the momentum of inertia~\citep{Lattimer_Prakash_2001},\footnote{
We take a fiducial NS mass and radius of $M_{\rm ns} = 1.4 \ M_\odot$ and $R_{\rm ns} = 12 \ {\rm km}$, respectively.}
%The rotation energy can be as large as $\approx 0.5 I \Omega_{\rm K}{}^2 \sim 7.6 \times 10^{52} \ \rm erg$, 
%where $\Omega_{\rm K} = (G M_{\rm ns}/R_{\rm ns}{}^3)^{1/2}$ is the Keplarian breakup velocity. 
%These parameters are fiducial for cold NS, and not sufficiently viable for the earliest phase of the proto-NS formation, 
%namely within the Kelvin-Helmholtz timescale, $t_{{\rm KH}, \nu} \sim 10\mbox{-}100 \ \rm s$, 
%where the proto-NS is more extended, $R_{\rm ns} \sim 30\mbox{-}40 \ \rm km$~\citep[e.g.,][]{Suwa_2014}.
and $P_{\rm i}$ is the initial spin period. 
Even cases with $P_{\rm i} \gtrsim 10 \ \rm ms$ have a free energy of ${\cal E}_{\rm rot, i} \sim 10^{50} \ \rm erg$, which is sufficient to power SN explosions. 

\begin{figure}
\centering
\includegraphics[width=90mm]{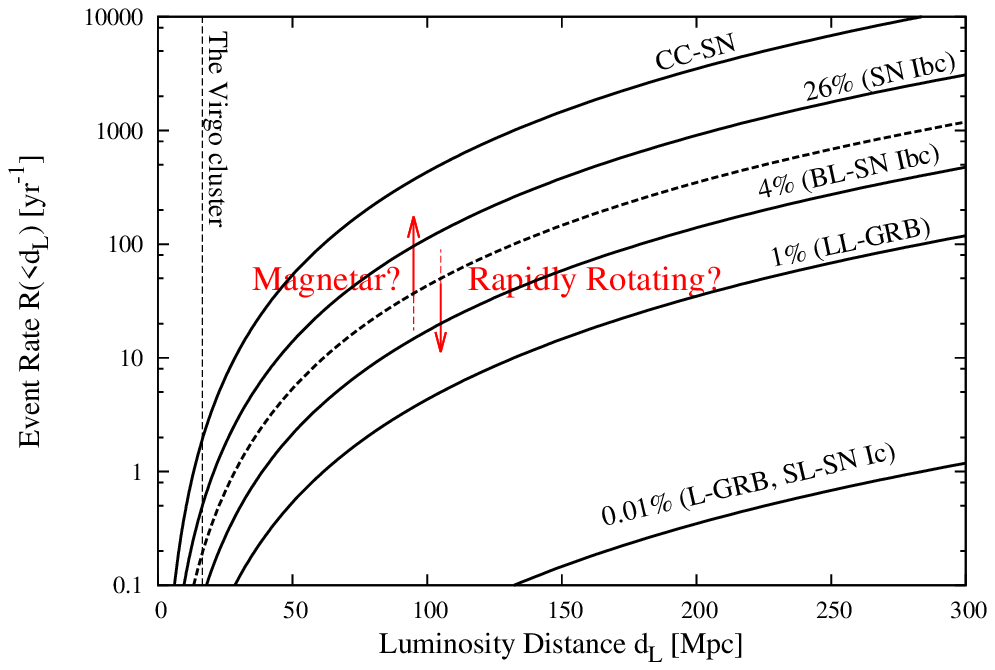}
\caption{
All-sky event rate of explosive phenomena potentially explained by pulsar-driven SN model. 
The core-collapse supernova (CC-SN) rate is obtained from \cite{Horiuchi_et_al_2011}.
We also refer to \cite{Smith_et_al_2011}, \cite{Guetta_and_Della_2007}, \cite{Wanderman_Piran_2010} and \cite{Quimby_et_al_2013} 
for the rates of supernova Ibc and the peculiar sub-class with broad-line features (BL-SN Ibc), low-luminosity GRB (LL-GRB), long GRB (L-GRB), 
and hydrogen-poor superluminous supernova (SL-SN Ic), respectively.
%The magnetar formation rate is from \cite{Keane_Kramer_2008}. 
The vertical dashed line shows the mean distance to the Virgo cluster ($= 16.5 \ \rm Mpc$).
}
\label{fig:rate}
\end{figure}

Formation of fast-spinning strongly magnetized proto-NSs may not be rare.  
%Observationally, massive O-type stars typically rotate with a equatorial velocity of $\sim 200 \ \rm km \ s^{-1}$ and beyond~\citep{Fukuda_1982}, 
%which is not much less than the brake-up limit. 
%Such massive stars with $\sim$ a solar metalicity will evolve into WRs after the significant mass loss. 
%Numerical simulations show that the spin period of proto-NS is sensitive to that of the pre-collapse iron core, and can typically range up from a few ms to a few $10 \ \rm ms$~\cite[e.g.,][]{Heger_et_al_2005,Ott_et_al_2006}. 
%However, angular-momentum transfer processes during the massive stellar evolution is still far from complete understanding. 
%The core may spin down significantly due to the magnetic braking~\citep[e.g.,][]{Meynet_et_al_2011}. 
%Also, a significant fraction of massive stars are in close binaries, and the binary interaction strongly affects the evolution of the angular momentum~\citep{Sana_et_al_2012}. 
%The spin period of proto-NS can be also inferred from the current spin-down characteristics and the remnant ages of Galactic pulsars. 
%The initial spin of relatively young pulsars e.g., the Crab pulsar, is estimated to be $\sim$ a few $10 \ \rm ms$.  
Population synthesis calculations of Galactic NS pulsars showed that the initial spin distribution is a Gaussian with a peak at $\sim 200\mbox{-}300 \ \rm ms$ 
and standard deviation of $\sim 100 \ \rm ms$~\citep{FaucherGigu_Kaspi_2006,Popov_et_al_2010}. 
If this applies even outside the Galaxy then the formation rate of proto-NSs with $P_{\rm i} \lesssim 30 \ \rm ms$ is:
\begin{equation}\label{eq:rapid_rot}
{\cal R}_{P_{\rm i} \ \lesssim 30 \ \rm ms} \sim 0.1 \times {\cal R}_{\rm CC-SN}. 
\end{equation}
Here, ${\cal R}_{\rm CC-SN} \sim 0.03^{+0.073}_{-0.026} \ \rm yr^{-1} {gal}^{-1}$ is the core-collapse SN rate~\citep[e.g.,][]{Adams_et_al_2013}. 
On the other hand, the formation rate of Galactic magnetar has been estimated from the observed spin-down rate and remnant age~\citep{Keane_Kramer_2008}: 
\begin{equation}\label{eq:magnetar}
{\cal R}_{\rm magnetar} \sim 0.1 \times {\cal R}_{\rm CC-SN}. 
\end{equation}
Although uncertainties in both Eq. (\ref{eq:rapid_rot}) and (\ref{eq:magnetar}) are fairly large, 
the formation rate of fast-spinning proto-NSs and Galactic magnetars is still consistent with the dynamo scenario. 
At this stage, there is no observational support on this interesting possibility~\citep[see, e.g.,][]{Vink_Kuiper_2006}. 

The progenitors of Galactic magnetars are considered to be very massive stars with $M_{\rm ZAMS} \gtrsim 30-40 M_\odot$ 
based on the fact that they are observed in young massive star clusters~\citep[e.g.,][]{Figer_et_al_2005,Gaensler_et_al_2005,Muno_et_al_2006,Bibby_et_al_2008,Davies_et_al_2009},  
and distributed in low Galactic latitudes~\citep{Olausen_Kaspi_2014}. 
Note that the fraction of massive stars with $M_{\rm ZAMS} \gtrsim 40 M_\odot$ is roughly $\sim 10 \%$ of that of  $M_{\rm ZAMS} \gtrsim 8 M_\odot$ given the Salpeter initial-mass function.
Such massive progenitors with approximately a solar metallicity are considered to evolve into Wolf-Rayet stars~(WRs), and end their lives as SE SNe~\citep[e.g.,][]{Heter_et_al_2003}.\footnote{
We note that a significant fraction of the observed SE SNe may be from close binary systems~\citep[e.g.,][]{Eldridge_et_al_2008,Smith_et_al_2011}.} 
%or, observationally, it has been suggested that the collapses can occur also in the luminous blue variable phases. 
%Note that if the progenitor is in a binary system, the pre-collapse structure can be altered by the mass and angular momentum transfer in the binary interaction. 
%Also, SNe associated with GRBs are so far all peculiar type Ibc. 
%If progenitors of magnetar are indeed as massive as above, associated SNe are unlikely type II-P whose progenitors are RSGs. 
%Note, however, that some observed type II-L SN can be well explained by the magnetar scenario~\citep{Kasen_Bildsten_2010}. 
%It is observationally indicated that the difference between SNe II-P and II-L is not only the amount of H envelope, 
%but also some other characteristics of progenitor and remnant compact object~\citep{Arcavi_et_al_2012}. 
The observed event rate of energetic SE-SNe and associated high-energy transients are relatively low:
\begin{equation}\label{eq:BL-Ibc}
{\cal R}_{\rm BL-SN \ Ibc} \sim 0.04 \times  {\cal R}_{\rm CC-SN}, 
\end{equation}
\begin{equation}
{\cal R}_{\rm LL-GRB} \sim 0.01 \times {\cal R}_{\rm CC-SN}, 
\end{equation}
\begin{equation}\label{eq:SL-SNrate}
{\cal R}_{\rm L-GRB} \sim {\cal R}_{\rm SL-SN \ Ic} \sim 10^{-4} \times {\cal R}_{\rm CC-SN},
\end{equation}
\citep{Guetta_and_Della_2007,Wanderman_Piran_2010,Smith_et_al_2011,Quimby_et_al_2013}. 
As for L-GRB, a jet beaming factor of $\sim 100$ is assumed~\citep[e.g.][]{Guetta_et_al_2005}. 
Even if these transients are powered by newborn magnetars, they explain only a minor fraction of the total magnetar abundance. 
On the other hand, ordinary SN Ibc can meet the magnetar formation rate as high as Eq. (\ref{eq:magnetar}):
\begin{equation}\label{eq:Ibc}
{\cal R}_{\rm Ibc} \sim 0.26 \times {\cal R}_{\rm CC-SN}. 
\end{equation}
Actually, e.g., \cite{Maeda_et_al_2007} proposed a newborn magnetar as a relevant energy source of a type Ibc SN 2005bf. 
In this regard, it is important to show in what parameter range ordinary SN Ibc is compatible with the pulsar-driven scenario and how to identify the underlying newborn pulsars.   

We should note that it has been pointed out that pulsar-driven models cannot reproduce the observed light curves of SNe Ibc and BL-SNe Ibc, 
in particular the late-time behavior $\gtrsim 100 \ \rm days$ after the explosion~\citep[e.g.,][]{Sollerman_et_al_2002,Inserra_et_al_2013}. 
In such a late phase, however, theoretical modeling of the optical light curve is still largely uncertain (see Sec. \ref{sec:late_phase} and Sec. \ref{sec:caveats}). 
In this paper, we focus on the optical light curves around the peak where the diffusion approximation is robust. 
On the other hand, in the late phase, the SN ejecta becomes almost transparent for nascent PWN emissions in hard X rays and gamma rays. 
They can be good probes of the properties of the underlying pulsar and one should take into account the fact that only a fraction of the energy can be converted into optical emission since high-energy emission escapes.

\section{Pulsar-Driven Supernova Scenarios}\label{sec:model}

\begin{figure}
\centering
\includegraphics[width=90mm]{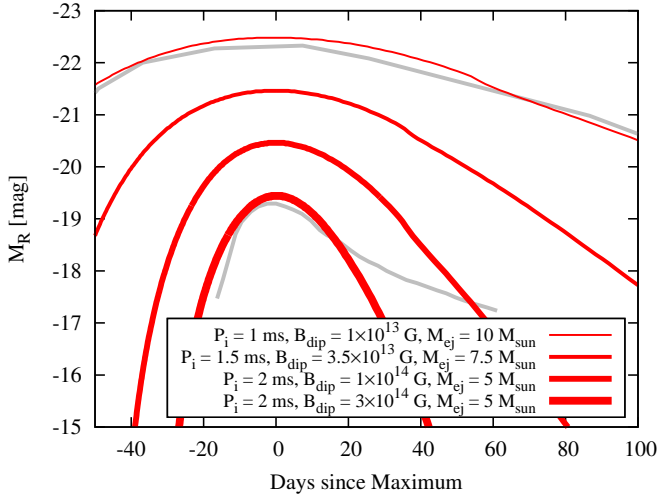}
\caption{
SN light curves of pulsar-driven SN models with ms rotation and different magnetic field strength. 
The upper and lower gray lines show the observed SL-SN-Ic PTF 09cnd and BL-SNe Ic 1998bw, respectively. 
}
\label{fig:sn_height}
\end{figure}

As popularly discussed in the literature~\citep[e.g.,][]{Ostriker_Gunn_1969,Thompson_et_al_2004,Woosley_2010, Kasen_Bildsten_2010,Wang_et_al_2015}, 
very bright SNe could be explained by the pulsar-driven SN model with $P_{\rm i}$ less than $\rm a \ few \ ms$. 
In this scenario, the peak luminosity of the pulsar-driven SN can be estimated as $L_{\rm sn}^{\rm psr} \approx {\cal E}_{\rm rot, i} \times [t_{\rm sd}^{\rm em}/(t_{\rm esc}^{\rm ej})^2]$~\citep{Kasen_Bildsten_2010}, or 
\begin{eqnarray}\label{eq:L_sn_psr}
L_{\rm sn}^{\rm psr} \sim 3 \times 10^{44} && \ \rm erg \ s^{-1} \  \left(\frac{B_{\rm dip}}{10^{14} \ \rm G}\right)^{-2} \left(\frac{M_{\rm ej}}{5 \ M_\odot}\right)^{-1} \notag \\
&& \times \left(\frac{V_{\rm ej}}{2 \times 10^9 \ \rm cm \ s^{-1}}\right) \left(\frac{K_{\rm T}}{0.2 \ \rm g^{-1} \ cm^2}\right)^{-1}. 
\end{eqnarray}
Here, 
\begin{equation}\label{eq:tsd_em}
t_{\rm sd}^{\rm em} \sim 0.4  \ {\rm days} \ \left(\frac{B_{\rm dip}}{10^{14} \ \rm G}\right)^{-2} \left(\frac{P_{\rm i}}{\rm ms} \right)^{2} 
\end{equation}
is the dipole spin-down timescale,\footnote{
As for the spin-down luminosity, we use a formula motivated by up-to-date MHD simulations, which give a factor $9/2$ larger value on average than the classical dipole formula (see Eq. \ref{eq:L_m_1}). 
As a result, $t_{\rm sd}^{\rm em}$ and $L_{\rm sn}^{\rm psr}$ become smaller by the same factor for a given $B_{\rm dip}$ and $P_{\rm i}$. This difference may affect the estimation of these parameters from observations.}
and  
%\begin{equation}\label{eq:L_m_2}
%L_{\rm em, i} \sim 1.6 \times 10^{47} \ {\rm erg \ s^{-1}} \ \left(\frac{B_{\rm dip}}{10^{14} \ \rm G}\right)^2 \left(\frac{P_{\rm i}}{\rm ms}\right)^{-4}  
%\end{equation}
%is the initial spindwon luminosity, 
\begin{eqnarray}
t_{\rm esc}^{\rm ej} \sim 20 \ {\rm days}  && \  \left(\frac{M_{\rm ej}}{5 \ M_\odot}\right)^{1/2} \left(\frac{V_{\rm ej}}{2 \times 10^9 \ \rm cm \ s^{-1}}\right)^{-1/2} \notag \\ 
&& \times \left(\frac{K_{\rm T}}{0.2 \ \rm g^{-1} \ cm^2}\right)^{1/2}.
\end{eqnarray}
is the photon diffusion time from the ejecta. 

In this work, we numerically calculate light curves of SNe driven by fast-spinning strongly magnetized newborn NSs embedded in SE progenitors.  
Details of the model description are given in the Appendix.  
We assume that the energy injection is caused by spherical winds rather than jets and both ${}^{56}$Ni decay and magnetized wind are taken into account as energy sources. 
The thermalization of the non-thermal emission is approximately taken into account, and the optical SN emission and early non-thermal nebular emission are obtained consistently.  
The effect of GW spin-down is incorporated in a simple parametric form.  The present model is based on \cite{Murase_et_al_2015} but with several refinements, e.g., including the effect of ${}^{56}$Ni decay.  
The simple model allows us to explore a wide parameter range; the initial spin of $P_{\rm i} = 1-30 \ \rm ms$, poloidal magnetic field of $B_{\rm dip} = 10^{13-15} \ \rm G$, SN ejecta mass of $M_{\rm ej} = 1-10 \ M_\odot$, 
${}^{56}$Ni mass of $M_{\rm {}^{56}Ni} = 0.05-1.0 \ M_\odot$, SN explosion energy of ${\cal E}_{\rm sn} = 10^{51-52} \ \rm erg$, and graybody opacity $K_{\rm T} = 0.05-0.2 \ \rm g^{-1} \ cm^{2}$. 
Note that $K_{\rm T} \sim 0.1$ and $0.2 \ \rm g^{-1} \ cm^{2}$ corresponds to electron scattering for singly ionized and fully ionized helium, respectively, 
and can be smaller for, e.g., a partially ionized C- or O-dominated ejecta. 

In Fig. \ref{fig:sn_height}, we show some light curve examples of the millisecond-pulsar-driven SN model. 
The thicker red lines correspond to larger magnetic fields. %with relatively large ejecta mass $M_{\rm ej} \geq 5 \ M_\odot$.   
The gray lines indicate the observed SL-SN-Ic PTF 09cnd~\citep{Quimby_et_al_2011,Gal-Yam_2012} and BL-SNe Ic 1998bw~\citep{Galama_et_al_1998}. 
Pulsar-driven SNe become as bright as SL-SNe with $B_{\rm dip} \lesssim 10^{14} \ \rm G$. 
In such cases, a significant fraction of the spin-down luminosity needs to be converted into SN radiation. 
%In particular, ASASSN-15lh~\citep{Dong_et_al_2015}, the most luminous SL-SN Ic ever discovered, is challenging in this regard. 
The pulsar-driven SN model with $B_{\rm dip} \sim 10^{13} \ \rm G$ and $P_{\rm i} \lesssim 1 \ \rm ms$ can reproduce the observed light curve of this event. 
%{\bf Say that almost all the spin-down energy needs to be converted into radiation for the model to work.}

We also consider the stronger case, where energy injection from fast-spinning NSs contributes to some BL-SNe Ibc and possibly ordinary SNe Ibc.  
For a fixed initial spin, the peak luminosity becomes smaller with a stronger magnetic field since $t_{\rm sd}^{\rm em}$ becomes smaller (see Fig. \ref{fig:sn_height} and Eqs. \ref{eq:L_sn_psr}-\ref{eq:tsd_em}).    
The physical reason is that the proto-NS spins down long before the photon diffusion time, and the injected energy by the pulsar wind is lost via adiabatic cooling. 
This means, on the other hand, that the injected energy is used for acceleration of the ejecta rather than SN radiation. 
Interestingly, as for $P_{\rm i}$ about $\rm a \ few \ ms$ and $B_{\rm dip} \gtrsim 5 \times 10^{14} \ \rm G$, 
the peak luminosity becomes $L_{\rm sn}^{\rm psr} \lesssim 10^{43} \ \rm erg \ s^{-1}$ and the mean ejecta velocity is $V_{\rm ej} \sim 20,000 \ \rm km \ s^{-1}$, 
which is compatible with the observed BL-SNe Ibc. 
An interesting possibility is that SL-SNe Ic and BL-SNe are connected sequences, and the main difference is the strength of the magnetic field. 

Although the pulsar-driven SN model can explain the peak light curve of BL-SNe Ibc, the radioactive decay of ${}^{56}$Ni has typically been considered as the main energy source, so as in the case of ordinary SNe Ibc.
The peak luminosity powered by the ${}^{56}$Ni decay can be roughly estimated as $L_{\rm sn}^{\rm {}^{56}Ni} \approx L_{\rm {}^{56}Ni} \times (t_{\rm {}^{56}Ni}/t_{\rm esc}^{\rm ej})^2$, or 
\begin{eqnarray}\label{eq:L_sn_ni}
L_{\rm sn}^{\rm {}^{56}Ni} \sim 4 \times 10^{42} && \ \rm erg \ s^{-1} \ \left(\frac{M_{\rm {}^{56}Ni}}{0.1 \ \rm M_\odot}\right) \left(\frac{M_{\rm ej}}{5 \ M_\odot}\right)^{-1} \notag \\
&& \times \left(\frac{V_{\rm ej}}{10^9 \ \rm cm \ s^{-1}}\right) \left(\frac{K_{\rm T}}{0.05 \ \rm g^{-1} \ cm^2}\right)^{-1}. 
\end{eqnarray}
On the other hand, the observed bolometric luminosities range from $\sim 10^{42-43} \ \rm erg \ s^{-1}$ for SN Ibc and $\sim 10^{43} \ \rm erg \ s^{-1}$ for BL-SN Ibc. 
The synthesized ${}^{56}$Ni masses are estimated to be $\sim 0.05-0.8 \ M_\odot$, although the uncertainties are large~\citep[e.g.,][]{Drout_et_al_2011,Lyman_et_al_2014}. 

\begin{figure}
\centering
\includegraphics[width=90mm]{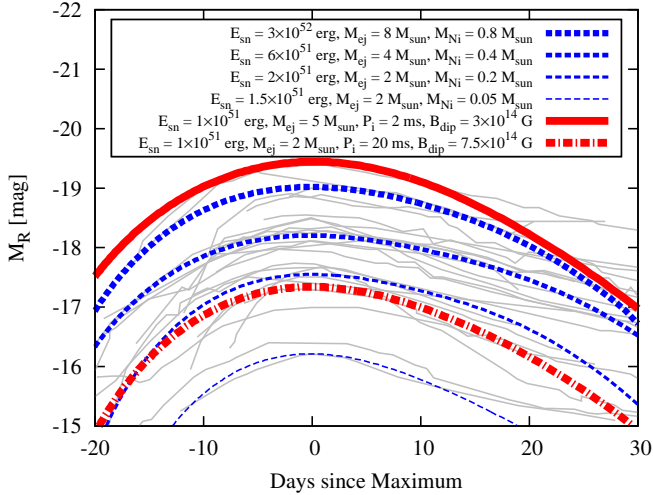}
\caption{
Comparison of observed SN Ibc and BL-SN Ibc (gray lines) and theoretical light curves. 
The red solid and dotted-dashed lines show wind-powered cases, whereas 
the blue dashed lines show ${}^{56}$Ni-powered cases. 
}
\label{fig:sn_degenerate}
\end{figure}

Fig. \ref{fig:sn_degenerate} shows several sample light curves.  
The blue dashed lines are the cases in which only ${}^{56}$Ni decay is considered.  
The gray lines are the observed light curves of SNe Ibc and BL-SN Ibc~\citep{Drout_et_al_2011}.  
Comparing Eqs. (\ref{eq:L_sn_psr}) and (\ref{eq:L_sn_ni}), one sees that the pulsar-driven model may also mimic SN light curves, 
with the flux as dim as that of observed SN Ibc by considering a relatively large magnetic fields, $B_{\rm dip} \gtrsim 5 \times 10^{15} \ \rm G$.  
Note that a relatively slow rotation of $P_{\rm i} \gtrsim 10 \ \rm ms$ better explains ordinary SNe Ibc; 
if the spin is faster, the SN ejecta is inevitably accelerated up to a high velocity and the SN becomes brighter (see Eq. \ref{eq:L_sn_psr}). 
At this stage, one could speculate that some of the BL-SNe Ibc and SNe Ibc are also connected sequences. 
Both can be driven or aided by newborn pulsars with a magnetar-class dipole field and the difference is the spin. 
%Both the ${}^{56}$Ni-powered model and the pulsar-driven SN model adequately describe the temporal behavior of the SN light-curves. 

\subsection{Optical Constraints on $P_{\rm i}$ and $B_{\rm dip}$}

\begin{figure}
\centering
\includegraphics[width=90mm]{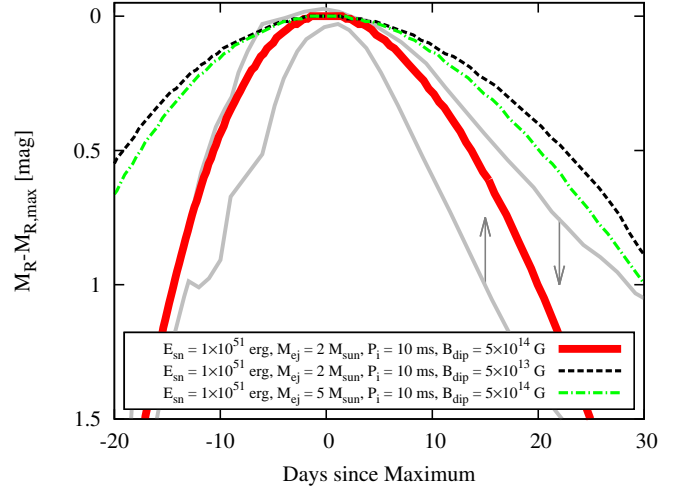}
\caption{
SN light curve of pulsar-driven SN models around the peak. 
The gray lines with arrows indicate the observed range of SNe Ibc and BL-SN Ibc. 
}
\label{fig:sn_width}
\end{figure}

In addition to the peak luminosities we discussed above, rising and decaying timescales of SN light curves can be used to constrain physical parameters of underlying proto-NSs. 
Fig. \ref{fig:sn_width} focuses on the raising and early decline of light curves.
The gray lines indicate the observed range of SN Ibc and BL-SN Ibc~\citep{Drout_et_al_2011}. 
The decline rate is in the range of $0.3 \lesssim M_{\rm R, 15} \lesssim 1$, where $M_{\rm R, 15}$ is $M_{\rm R} - M_{\rm R, max}$ at $15$ days after the peak. 
The thick solid red line shows a pulsar-driven case broadly consistent with the observed SNe Ibc.
The evolution of a light curve becomes wider when the poloidal field is smaller (dash line) because the energy injection rate declines more slowly.
Also, a larger ejecta mass case (dotted dash line) gives a slow light curve because the photon diffusion time becomes longer.  

\begin{figure}
\centering
\includegraphics[width=90mm]{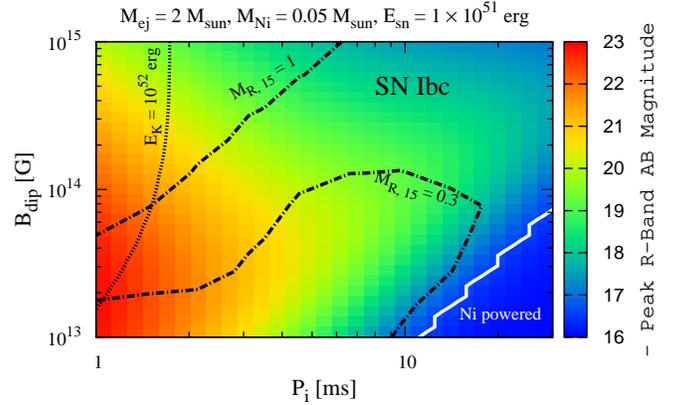}
\caption{ 
Contour plot showing properties of SN counterpart of fast-spinning strongly magnetized proto-NS formed 
with $M_{\rm ej} = 2 \ M_\odot$, $M_{\rm {}^{56}Ni} = 0.05 \ M_\odot$, ${\cal E}_{\rm sn} = 1 \times 10^{51} \ \rm erg$ and $K_{\rm T} = 0.05 \ \rm g^{-1} \ cm^2$. 
The color with solid lines shows the peak absolute magnitude, 
the dotted-dashed lines show the decline rate of the light curve, $M_{\rm R, 15} = 0.3$ and $1$, and 
the dotted line shows the contour of ${\cal E}_{\rm K} = 1 \times 10^{52} \ \rm erg$. 
The parameter region broadly consistent with the observed SN Ibc is indicated. 
}
\label{fig:sn_mej2}
\end{figure}

\begin{figure}
\centering
\includegraphics[width=90mm]{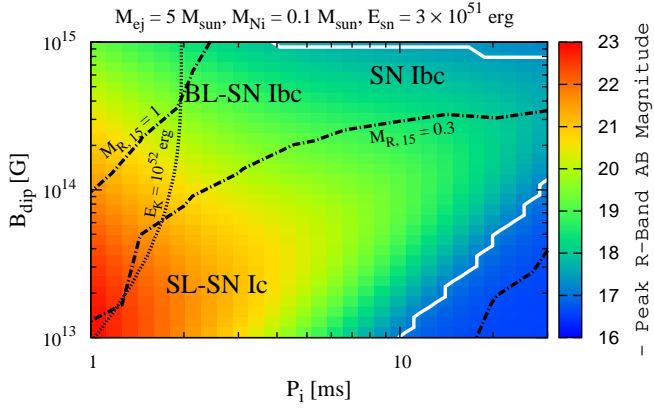}
\caption{ 
Same as Fig. \ref{fig:sn_mej2}, but with  $M_{\rm ej} = 5 \ M_\odot$, $M_{\rm {}^{56}Ni} = 0.1 \ M_\odot$, and ${\cal E}_{\rm sn} = 3 \times 10^{51} \ \rm erg$. 
Parameter regions broadly consistent with the observed SN Ibc, BL-SN Ibc, and SL-SN Ic are indicated. 
}
\label{fig:sn_mej5}
\end{figure}

In Figs. \ref{fig:sn_mej2} and \ref{fig:sn_mej5}, we show in what parameter range the pulsar-driven SN model can explain the observed optical emission from SE SNe. 
Fig. \ref{fig:sn_mej2}(\ref{fig:sn_mej5}) corresponds to relatively low (high) ejecta mass, $M_{\rm ej} = 2 \ M_\odot$ ($5 \ M_\odot$). 
In both cases, the ${}^{56}$Ni mass and SN explosion energy is moderate and the SN emission is predominantly powered by the magnetized wind except for the bottom right conner of the panels.
The boundary of the Ni dominated region is shown by the solid white line. 
SNe Ibc with $M_{\rm R, max} \sim -(17-18)$ and $M_{\rm R, 15} \sim 0.3-1$ can be explained by the pulsar-driven SN model in the top right conner of the panels, 
$P_{\rm i} \gtrsim 10 \ \rm  ms$ and $B_{\rm dip} \gtrsim 5 \times 10^{14} \ \rm G$. 
Note that proto-NSs with relatively weak poloidal fields cannot hide in SNe Ibc since the light curves become slower than the observed ones.   
BL-SNe Ibc with $M_{\rm R, max} \sim -(18-19)$ and $V_{\rm ej} \sim 20,000 \ \rm km \ s^{-1}$ can be explained by a larger-mass case, $M_{\rm ej} \gtrsim 5 \ \rm M_\odot$, with 
$P_{\rm i}$ about $\rm a \ few \  ms$ and $B_{\rm dip} \gtrsim  5 \times 10^{14} \ \rm G$, in which the kinetic energy is also mainly provided by the magnetized wind. 
SL-SNe Ic also prefer relatively large ejecta mass cases since their light curves are relatively slow, the decrease in magnitude 40 days after peak is $< 1.5$; \citep{Quimby_et_al_2013}.%$M_{\rm R, 40} \sim 1.0$. 
The best fitting parameter range is $P_{\rm i}$ less than $\rm a \ few \ ms$ and $B_{\rm dip} \gtrsim 10^{13} \ \rm G$.  

The possibility that a significant fraction of SE-SNe are driven by nascent pulsars is interesting in view of the connection among GRBs, SL-SNe and BL-SNe~\citep[see also][]{Metzger_et_al_2015}. 
It is also of interest in view of the connection to Galactic magnetars in the dynamo hypothesis.    

\subsection{Late-time behavior}\label{sec:late_phase}
As shown above, peak optical light curves of SNe Ibc and BL-SNe Ibc can be broadly explained by the pulsar-driven model with the appropriate choice of $P_{\rm i}$ and $B_{\rm dip}$.  
On the other hand, these SNe have been considered to be mainly powered by ${}^{56}$Ni decay. 
The parameter degeneracy between $P_{\rm i}$, $B_{\rm p}$, and $M_{\rm {}^{56}Ni}$ cannot be solved only from the peak optical light curves. 
One promising way is to use late-time spectroscopy. 
Indeed, in some cases, the ${}^{56}$Ni masses are independently determined by observing Fe line emissions in the Co decay phase ($\gtrsim 100 \ \rm days$), 
and are consistent with the values obtained from the peak optical light curves. 
However, such observations are challenging since the line emissions are typically very faint. 
Also, there are still significant uncertainties in the line transfer calculation. 
Another possible way to solve the parameter degeneracy is to use the late-time optical photometry from SNe, 
which can provide an independent constraint on the ${}^{56}$Ni mass. 
However, it is known that the late-time light curves of SE SNe are heterogeneous 
and difficult to fit consistently with the optical peak using a simple ${}^{56}$Ni-decay model~\citep[e.g.,][]{Wheeler_et_al_2015}.
Also, pulsar-driven models could reproduce the light curves.
Note that our simple model of calculating optical light curves becomes less reliable after the late decline phase or early nebular phase $\sim 20 \ \rm days$ after the peaks. 
More detailed theoretical calculations of late-time optical emission are necessary. 

\section{Multi-Messenger Tests}
Because of additional parameters in the pulsar-driven SN model, optical light curves alone may not be used to distinguish the model from the other competing models.  
Multi-messenger approaches are useful to break parameter degeneracies, to test the pulsar-driven scenario for SE SNe from ordinary SN Ibc to BL-SN Ibc and SLSN Ic and also the Galactic magnetar connection to SE SNe.  
A unique signature of newborn pulsar engines is the PWN emission 
in X rays~\citep[e.g.,][]{Perna_et_al_2008,Metzger_et_al_2013,Murase_et_al_2015} and gamma rays~\citep{Kotera_et_al_2013, Murase_et_al_2015}. 
Although the dissipation mechanism of the magnetized wind is still controversial, a most likely outcome is an injection of ultra-relativistic electrons, 
which triggers leptonic pair cascades mediated via synchrotron emission and (inverse) Compton scattering. 
The synthesized nebular emissions are entirely down-scattered into the thermal bath in the earlier phase of the ejecta expansion, 
but start to escape the ejecta at a later time.  
By observing such broadband nebular emissions in soft X-ray, hard X-ray, and gamma-ray bands, it is possible to put independent constraints on the physical parameters of underlying NSs.  
Such signals can also probe the particle acceleration in embryonic PWNe. 

Moreover, fast-spinning strongly magnetized proto-NSs are possible sources of new messengers. 
In general, fast-spinning proto-NSs are unstable to non-axisymmetric perturbations and can evolve into a plausible configuration for emitting GWs~\citep[e.g.,][]{Kokkotas_2008,Bartos_et_al_2013}.  
The GW frequency is $f \sim 100 \ {\rm Hz} - 1 \ {\rm kHz}$, which coincides with the target frequency range of ground-based interferometers. 
In principle, the detection of such GWs can be used to determine physical parameters of newborn pulsars, e.g., the rotation period and deformation rate. 
%The details of the GW form depend on how much and for how long the nascent NS is deformed, which couple with highly uncertain physical properties,  
%e.g., the equation of state, viscosity, thermal evolution via neutrino cooling, magnetic field and its amplification.  
Neutrinos are also a powerful messengers. 
In addition to multi-MeV thermal neutrinos from proto-NSs, some hadron acceleration processes can occur in the magnetized wind or jet, 
and the energy dissipation results in GeV to EeV neutrino emissions~\citep{Murase_et_al_2014,Murase_et_al_2009,Fang_et_al_2014,Lemoine_et_al_2015}. 
Such high-energy neutrinos can be a probe of the physics in strongly magnetized winds. 

\subsection{High-energy X-ray and gamma-ray emission}\label{sec:X-ray}
Non-thermal emission from PWNe can probe underlying newborn pulsar engines. 
The injection spectrum is a hard power law, $dN_\gamma/dE_\gamma \propto E_\gamma^{-s}$ with $s \sim 1.5-2.5$ from soft X rays to GeV-TeV gamma rays \citep[][see also Sec.\ref{sec:em}]{Murase_et_al_2015}. 
The light curve depends on the spin-down of the underlying NS.%, thus can be a probe of its physical parameters. 

Here, we focus on the hard X-ray counterpart, where the Compton scattering is the main interaction process inside the SN ejecta and our theoretical calculation is most robust. 
We discuss the detectability using {\it NuSTAR}~\citep{Harrison_et_al_2013}, which operates in the band from 3 to 79 keV. 
Hard X rays can be also produced in the ${}^{56}$Ni-powered model;  
the gamma-rays produced by the ${}^{56}$Ni decay into ${}^{56}$Co and ${}^{56}$Fe with $\gtrsim \rm MeV$ are successively Compton-scattered down to lower energies. 
However, such hard X rays begin to be suppressed once the SN ejecta becomes Compton thin, while the PWN emission rises at the same moment.  
Moreover, the spectrum from Ni decay have a lower energy cutoff at $\sim 100 \ \rm keV$~\citep[e.g.,][]{Maeda_2006}, and can be distinguishable from the PWN spectrum.    

We should note that the PWN emission in other energy bands can be a useful counterpart too. 
If the energy injection by the pulsar wind is large enough and the ejecta mass is relatively small, the ionization break occurs and soft-X-ray can be observed~\citep{Metzger_et_al_2013}.
Such a situation is promising in SLSN-Ic, but not guaranteed in BL-SN Ibc and SN Ibc. 
Also, the GeV gamma-ray counterpart can be better than X rays as a probe since it does not depend on the ionization state.
Gamma-ray detection is possible for nearby objects up to $\lesssim$ 10 \ \rm Mpc as shown in \cite{Murase_et_al_2015}.  

\begin{figure}
\centering
\includegraphics[width=90mm]{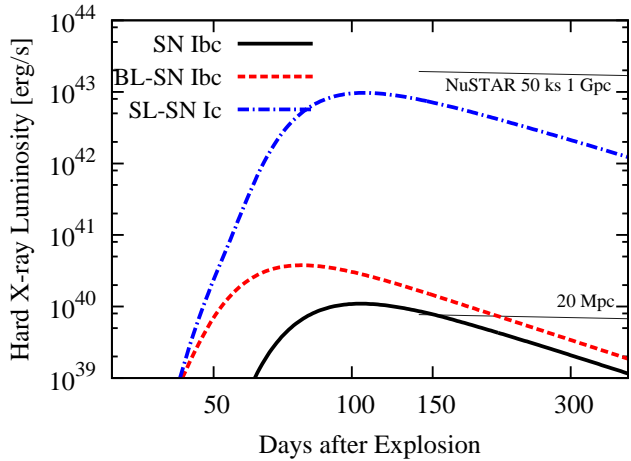}
\caption{
Hard X-ray ($30-80 \ \rm keV$) light curves of pulsar-driven SN model with typical parameters for SN Ibc, BL-SN Ibc, and SL-SN Ic labeled in Figs. \ref{fig:sn_mej2} and \ref{fig:sn_mej5}.  
The $3 \sigma$ detection threshold by 50 ks observation using {\it NuSTAR} from $20 \ \rm Mpc$ and $300 \ \rm Mpc$ is shown.
% corresponding to the observed flux of $\sim 8.9 \times 10^{-8} \ \rm ph \ cm^{-2} \ s^{-1} \ keV^{-1}$  
}
\label{fig:x_lc}
\end{figure}

Fig. \ref{fig:x_lc} shows light curve examples of pulsar-driven SN model in the hard-X-ray band ($30-80 \ \rm keV$). 
The thick, dash, and dotted-dashed lines correspond to the cases for which the SN counterpart is consistent with SN Ibc, BL-SN Ibc, and SL-SN Ic labeled in Figs. \ref{fig:sn_mej2} and \ref{fig:sn_mej5}, respectively. 
We also show the $3 \sigma$ detection threshold using {\it NuSTAR} with $50 \ \rm ks$ observation.  
The hard-X-ray counterpart can be detectable for SN Ibc and BL-SN Ibc at $\lesssim 30-50 \ \rm Mpc$ and for SL-SN Ic at $\lesssim 1 \ \rm Gpc$ ($z \lesssim 0.2$). 
The anticipated detectable event rate is $\sim 1 \ \rm yr^{-1} \ sky^{-1}$ for SN Ibc, BL-SN Ibc, and SLSN-Ic (see Fig. \ref{fig:rate}).
A follow-up observation needs to be undertaken $\sim 50-100 \ \rm days$ after the explosion. 
%For example, the hard-X-ray counterpart of ASASSN-15lh~\citep{Dong_et_al_2015} at $z \sim 0.2326$ could be detectable with a sufficiently long observation time, $\gtrsim 100 \ \rm ks$. 

The raising time of the hard X-ray counterpart can be roughly estimated from the condition $K_{\rm comp} \tau_{\rm comp}{}^2 \approx 1$, which gives  
\begin{equation}
t_{\rm X} \sim 80 \ \rm days \ \left(\frac{M_{\rm ej}}{5 \ M_\odot}\right)^{1/2} \left(\frac{V_{\rm ej}}{10^9 \ \rm cm \ s^{-1}}\right)^{-1} \left(\frac{E_\gamma}{50 \ \rm keV}\right)^{1/4}, 
\end{equation}
In the above estimate, we approximate the inelasticity of Compton scattering as $K_{\rm comp} \approx E_\gamma/m_{\rm e} c^2  (\ll 1)$. 
At $t > t_{\rm x}$, the PWN emission can be directly observed, i.e., $f_{\rm esc} \approx 1$.  
Here, $f_{\rm esc}(E_\gamma)$ is the fraction of the PWN emission escaping from the SN ejecta (see Eq. \ref{eq:f_esc}).
The peak luminosity is roughly given by $L_{\rm x} \approx L_{\rm em, i}/{\cal R}_{\rm b} \times (t_{\rm x}/t_{\rm sd}^{\rm em})^{-2}$, 
where the bolometric factor becomes ${\cal R}_{\rm b} \sim 10-20$ around the hard X-ray raising time. 
%\begin{equation}
%L_{\rm x} \approx \frac{L_{\rm em, i}}{{\cal R_{\rm b}}} \left(\frac{t_{\rm x}}{t_{\rm sd}^{\rm em}}\right)^{-2}.
%\end{equation}
From Eq. (\ref{eq:L_sn_psr}), the ratio between the SN and hard X-ray luminosity is given by 
\begin{eqnarray}
\frac{L_{\rm x}}{L_{\rm sn}^{\rm psr}} &\approx& \frac{1}{{\cal R_{\rm b}}K_{\rm comp}^{1/2}} \frac{V_{\rm ej}}{c}  \notag \\
&\sim& 0.007 \left(\frac{{\cal R}_{\rm b}}{15}\right)^{-1} \left(\frac{V_{\rm ej}}{10^9 \ \rm cm \ s^{-1}}\right) \left(\frac{E_\gamma}{50 \ \rm keV}\right)^{-1/2}.
\end{eqnarray}
These results are consistent with more detailed calculations by \cite{Murase_et_al_2015}. 

\begin{figure}
\centering
\includegraphics[width=90mm]{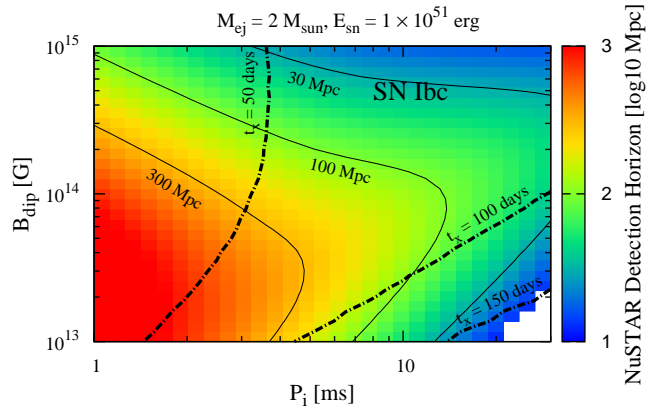}
\caption{
Contour plot showing properties of hard X-ray counterpart of fast-spinning strongly magnetized proto-NS formation 
with $M_{\rm ej} = 2 \ M_\odot$ and ${\cal E}_{\rm sn} = 1 \times 10^{51} \ \rm erg$. 
The color with solid lines corresponds to the $3 \sigma$ detection horizon using {\it NuSTAR} with $50 \ \rm ks$ observations at around the peak time shown in the dotted-dashed lines. 
}
\label{fig:x_mej2}
\end{figure}
 
\begin{figure}
\centering
\includegraphics[width=90mm]{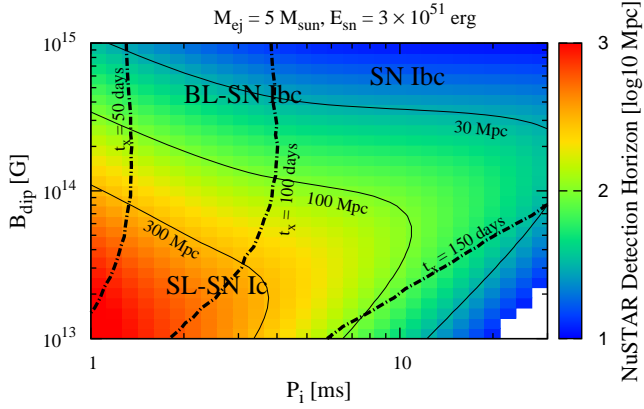}
\caption{ 
Same as Fig. \ref{fig:x_mej2}, but with  $M_{\rm ej} = 5 \ M_\odot$ and ${\cal E}_{\rm sn} = 3 \times 10^{51} \ \rm erg$. 
}
\label{fig:x_mej5}
\end{figure}

The detectability of the hard X-ray counterpart is shown for more general cases in Figs. \ref{fig:x_mej2} and \ref{fig:x_mej5}, for which we use the same parameter set as in Figs. \ref{fig:sn_mej2} and \ref{fig:sn_mej5} . 
The color contour with solid lines shows the detection horizon using {\it NuSTAR}, and the dotted-dashed lines show the emission raising time. 
The hard X-ray counterpart is a promising signature of identifying newborn pulsar engines. 

\subsection{Gravitational wave emission}
In general, fast-spinning proto-NSs are unstable to non-axisymmetric perturbations and can evolve into a plausible configuration for emitting GWs~\citep[e.g.,][]{Kokkotas_2008,Bartos_et_al_2013}.  
If the energy loss through the GWs is significant, then it might suppress the electromagnetic counterparts.  
Thus, it is useful to consistently model GW spin-down together with the electromagnetic spin-down.  
We focus on the GW emission due to magnetically deformed rotation~\citep{Cutler_2002,Stella_et_al_2005,DallOsso_et_al_2009}, which is an interesting channel especially in terms of magnetar formation. 
A proto-NS with a strong inner toroidal field $B_{\rm t}$ is deformed by the magnetic pinch effect. 
In general, the axis of the deformation is different from that of rotation and the proto-NS starts to precess.
The tilt angle of the precession increases secularly due to the bulk viscosity and the proto-NS evolves into a non-axisymmetric rotating body, which is a plausible configuration for the GW emission~(see the Appendix). 
The GW form is parameterized by $P_{\rm i}$, $B_{\rm p}$ and $B_{\rm t}$ (or, the deformation rate, $\varepsilon_{\rm G}$). 

First, let us argue the detectability of the GW counterpart under the competition with the electromagnetic spin-down.   
For a given spin-down timescale, i.e., $P(t)$ and GW luminosity $L_{\rm gw}(t)$, %, which is obtained by solving Eq. (\ref{eq:L}),  
the signal-to-noise ratio~(S/N) of the expected GW averaged over all possible orientations of source and detector can be estimated as~\citep{Owen_Lindblom_2002}; 
\begin{equation}\label{eq:s/n}
\left(\frac{S}{N}\right)^2 \approx -\frac{4G}{10 \pi c^3 D^2} \int \frac{dJ}{f S_{\rm h}(f)}.
\end{equation}
Here, $f = 2/P$ is the GW frequency, $dJ \equiv dL_{\rm gw}/(2 \pi/ P)$, and $S_{\rm h}(f)$ is the one-sided power spectral density of detector noise.  
We note that the $(S/N)$ calculated from Eq. (\ref{eq:s/n}) is roughly equal to that obtained by the matched-filtering analysis.
If the excess-power search is implemented, which is more appropriate for this type of GW, 
the anticipated $(S/N)$ would become smaller by a factor of a few to $\sim 10$~\citep{Thrane_et_al_2011,Piro_Thrane_2012}.
We use an anticipated sensitivity curve of Advanced LIGO~\footnote{https://dcc.ligo.org/LIGO-T0900288/public}, 
%We assume an observation time of $300$ days.  
which is for the optimal direction of the detector and the angle-averaged sensitivity is smaller by a factor of $\sim 2/3$. 
On the other hand, by combing other detectors, e.g., Advanced Virgo and KAGRA, the sensitivity effectively increases at most by a factor of $\sim \sqrt{3}$.

\begin{figure}
\centering
\includegraphics[width=90mm]{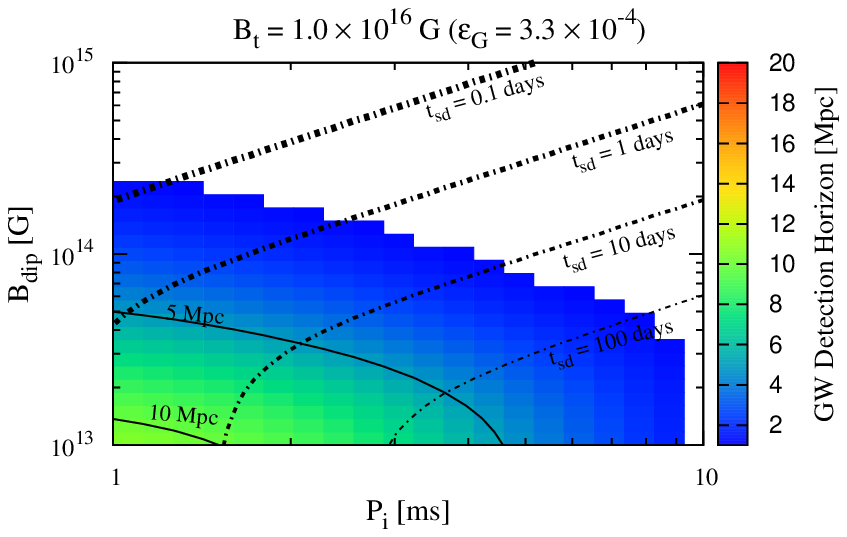}
\includegraphics[width=90mm]{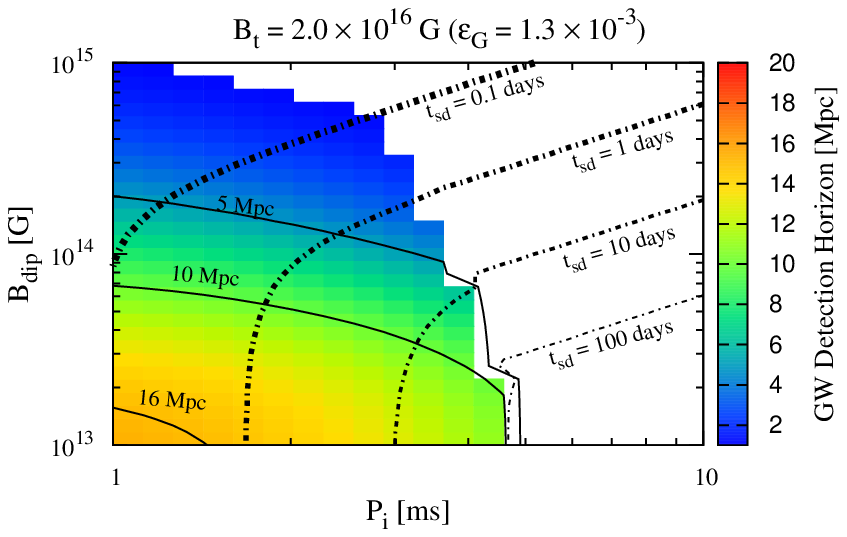}
\includegraphics[width=90mm]{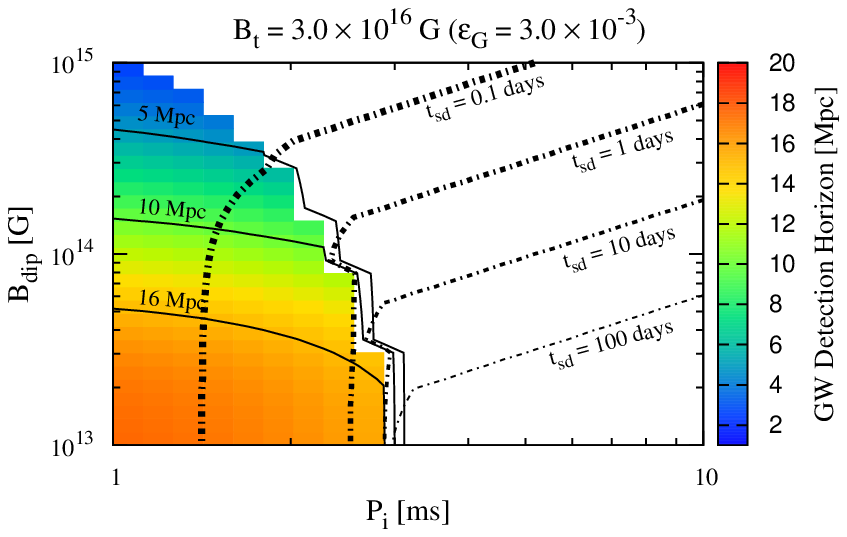}
\caption{
Detectability of the GW counterpart of magnetically deformed proto-NS. 
The solid-line contour with color shows the detection horizon of  $d_{\rm L} = 5, 10, 15$, and $20$ Mpc with $({\rm S/N}) = 8$ by using Advanced LIGO. 
The dotted-dash-line contour shows the spin-down timescale of the proto-NS, $t_{\rm sd} = 0.1, 1, 10$, and $100 \ {\rm days}$. 
}
\label{fig:gw}
\end{figure}

Fig. \ref{fig:gw} shows the detection horizon of the GW counterpart.  
Each panel shows a different deformation rate; $B_{\rm t} = 1.0 \times 10^{16} \ \rm G$ ($\epsilon_{\rm G} = 3.3 \times 10^{-4}$; top),  
$B_{\rm t} = 2.0 \times 10^{16} \ \rm G$ ($\epsilon_{\rm G} = 1.3 \times 10^{-3}$; middle), and $B_{\rm t} = 3.0 \times 10^{16} \ \rm G$ ($\epsilon_{\rm G} = 3.3 \times 10^{-3}$; bottom). 
Such a deformation rate is recently inferred for some Galactic magnetars from the X-ray timing observation~\citep{Makishima_et_al_2014,Makishima_et_al_2015}. 
The solid-line contour shows $d_{\rm L} = 5, 10, 15$, and $20$ Mpc with $({\rm S/N}) = 8$, which is the standard threshold value for compact binary mergers~\citep{LIGO_Virgo_2013}.
The dotted-dashed line contour shows the spin-down timescale of the proto-NS, $t_{\rm sd} \equiv (1/t_{\rm sd}^{\rm em} + 1/t_{\rm sd}^{\rm gw})^{-1} = 0.1, 1, 10$, and $100 \ {\rm days}$. 
In Fig. \ref{fig:gw} we shut off the GW spindwon for a larger toroidal field following Eq. (\ref{eq:dump}).  
%The initial spin-down power via the GW emission can be estimated as   
%\begin{equation}\label{eq:L_gw_2}
%L_{\rm gw, i} \sim 1.7 \times 10^{48} \ \epsilon_{\rm G, -3}{}^2 P_{\rm i, -3}{}^{-6} \ \rm erg \ s^{-1}, 
%\end{equation}
The spin-down timescale via GW emission can be roughly given by  
\begin{equation}\label{eq:tsd_gw}
t_{\rm sd}^{\rm gw} \sim 0.2 \ {\rm days} \ \left(\frac{\epsilon_{\rm G}}{10^{-3}}\right)^{-2} \left(\frac{P_{\rm i}}{\rm ms}\right)^{4} 
\end{equation}
From Eqs. (\ref{eq:tsd_em}) and (\ref{eq:tsd_gw}), the GW spin-down dominates when 
\begin{equation}
\epsilon_{\rm G} \gtrsim 3 \times 10^{-4} \ \left(\frac{B_{\rm dip}}{10^{14} \ \rm G}\right) \left(\frac{P_{\rm i}}{\rm ms}\right). 
\end{equation}
%One can see that $t_{\rm sd} \propto B_{\rm dip}{}^{-2} P_{\rm i}{}^2$ when the magnetic braking dominates whereas $t_{\rm sd} \propto P_{\rm 0}{}^4$ when the GW spin-down dominates (Eq. \ref{eq:tsd}). 
In general, the (S/N) becomes larger for a smaller dipole field because the competitive electromagnetic spin-down becomes irrelevant and for a faster rotation because the intrinsic energy budget becomes larger.
In principle, the GW can be detectable up to the Virgo cluster, $d_{\rm L} = 16.5 \ \rm Mpc$ 
for $P_{\rm i}$ less than $\rm a \ few \ ms$, $B_{\rm dip}$ less than ${\rm a \ few} \times 10^{13} \ \rm G$, and $B_{\rm t} \gtrsim 2 \times 10^{16} \ \rm G$ ($\epsilon_{\rm G} \gtrsim 10^{-3}$).  

With a sufficiently large (S/N), physical parameters like $P_{\rm 0}$ and $B_{\rm t}$ (or $\epsilon_{\rm G}$) can be determined from the GWs.  
The determination accuracies can be at most $\Delta P_{\rm i}/ P_{\rm i} \approx (S/N)^{-1}/\sqrt{N_{\rm cyc}}$ and $\Delta B_{\rm t}/B_{\rm t} \approx \Delta \epsilon_{\rm G}/\epsilon_{\rm G} \approx (S/N)^{-1}$.   
Here, $\sqrt{N_{\rm cyc}} \approx \sqrt{2t_{\rm sd}/ P_{\rm i}} \sim 6 \times 10^3 (\epsilon_{\rm G}/10^{-3})^{-1} (P_{\rm i}/{\rm ms})^{3/2}$ corresponds to the number of GW cycles in the spin-down timescale.  
Thus, if this type of GW is detected, the rotation period of proto-NS could be determined with a sufficient accuracy.  
%Note that the typical GW spin-down timescale can become $\sim 10 \ \rm days$ for $B_{\rm t} \lesssim 10^{16} \ \rm G$, thus it requires a relatively long-term data analysis~\citep{Thrane_et_al_2015}.  

\begin{figure}
\centering
\includegraphics[width=90mm]{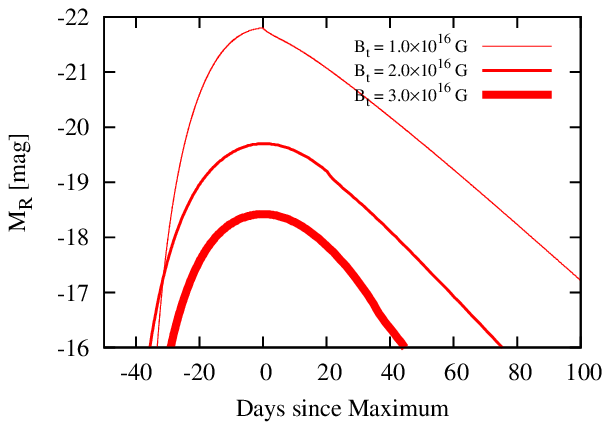}
\caption{
SN light curves of pulsar-driven SN models with millisecond rotation and different toroidal magnetic field strength, $B_{\rm t} = (1-3) \times 10^{16} \ \rm G$. 
We fix other parameters as $B_{\rm dip} = 2 \times 10^{13} \ \rm G$, $P_{\rm i} = 1 \ \rm ms$, $M_{\rm ej} = 2 \ M_\odot$, and $K_{\rm T} = 0.05 \ \rm g^{-1} \ cm^2$. 
}
\label{fig:sn_height_gw}
\end{figure}

\begin{figure}
\centering
\includegraphics[width=90mm]{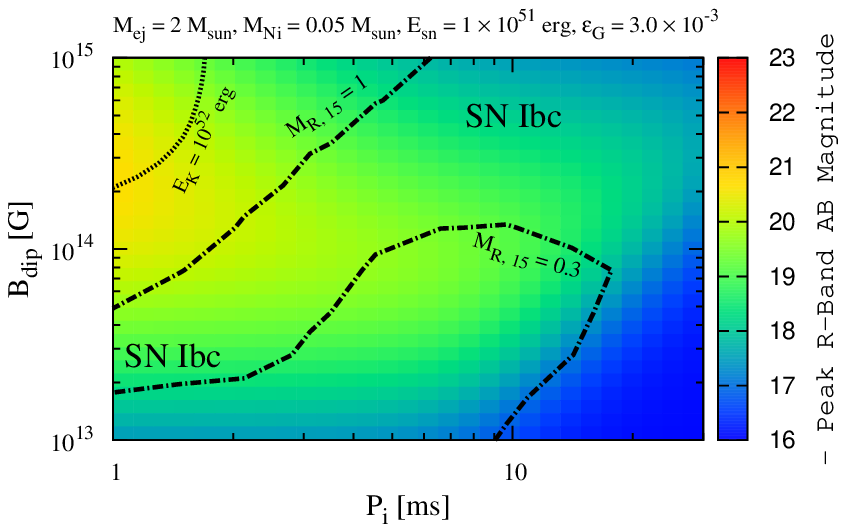}
\caption{ 
Contour plot showing properties of the SN counterpart of fast-spinning strongly magnetized proto-NS formation 
with $M_{\rm ej} = 2 \ M_\odot$, $M_{\rm {}^{56}Ni} = 0.05 \ M_\odot$, ${\cal E}_{\rm sn} = 1 \times 10^{51} \ \rm erg$, and $\epsilon_{\rm G} = 3.0 \times 10^{-3}$. 
The color with solid lines shows the peak absolute magnitude, 
the dotted-dashed lines show the decline rate of the light curve, $M_{\rm R, 15} = 0.3$ and $1.0$, and 
the dotted line shows the contour of ${\cal E}_{\rm K} = 1 \times 10^{52} \ \rm erg$. 
}
\label{fig:sn_mej2_bt3e16}
\end{figure}

\begin{figure}
\centering
\includegraphics[width=90mm]{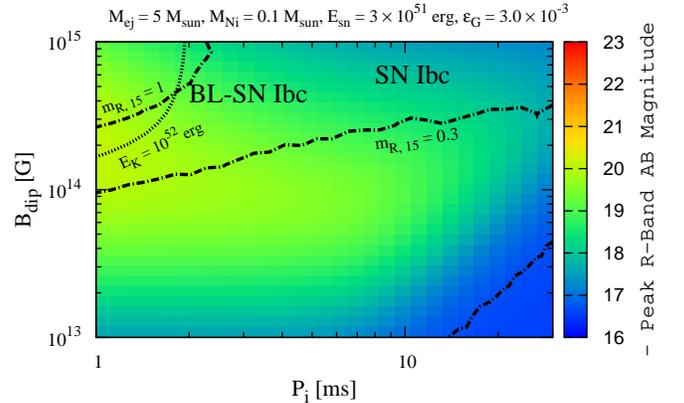}
\caption{ 
Same as Fig. \ref{fig:sn_mej2}, but with  $M_{\rm ej} = 5 \ M_\odot$, $M_{\rm {}^{56}Ni} = 0.1 \ M_\odot$, ${\cal E}_{\rm sn} = 3 \times 10^{51} \ \rm erg$.  
}
\label{fig:sn_mej5_bt3e16}
\end{figure}

Next, let us discuss the effect of GW spin-down on the electromagnetic counterpart. 
Fig. \ref{fig:sn_height_gw} shows several sample light curves of the pulsar-driven SN model with millisecond rotation and different toroidal magnetic field strength.
We set $B_{\rm dip} = 2 \times 10^{13} \ \rm G$, $P_{\rm i} = 1 \ \rm ms$, $M_{\rm ej} = 2 \ M_\odot$, and $K_{\rm T} = 0.05 \ \rm g^{-1} \ cm^2$. 
The SN emission becomes dimmer for a stronger GW spin-down, from the SL-SN class to the ordinary SN Ibc class. 
A broader parameter region is investigated in Figs. \ref{fig:sn_mej2_bt3e16} and \ref{fig:sn_mej5_bt3e16}, 
where we assume the same parameter set as in Figs. \ref{fig:sn_mej2} and \ref{fig:sn_mej5} except for  $B_{\rm t} = 3.0 \times 10^{16} \ \rm G$ ($\epsilon_{\rm G} = 3.0 \times 10^{-3}$).  
Comparing with Figs. \ref{fig:sn_mej2} and \ref{fig:sn_mej5}, the peak magnitude becomes significantly smaller in the parameter region where the GW spin-down is relevant. 
As a result, for a relatively small ejecta mass case (Fig. \ref{fig:sn_mej2_bt3e16}), the bottom left conner ($P_{\rm i}$ about $\rm a \ few \ ms$ 
and $B_{\rm dip}$ less than $\rm a \ few \ \times 10^{13} \ \rm G$) becomes consistent with ordinary SN Ibc.  
As for a relatively large ejecta mass case (Fig. \ref{fig:sn_mej5_bt3e16}), the parameter region compatible to SL-SN Ic disappears due to the GW spin-down.
For a larger poloidal magnetic fields, $B_{\rm dip} \gtrsim 10^{14} \ \rm G$, the effect of GW spin-down is not noticeable.  

\begin{figure}
\centering
\includegraphics[width=90mm]{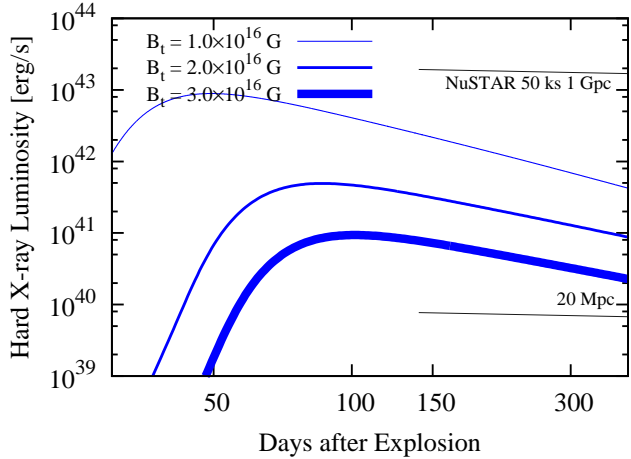}
\caption{
Hard X-ray ($30-80 \ \rm keV$) light curves of pulsar-driven SN models with millisecond rotation and different toroidal magnetic field strength, $B_{\rm t} = (1-3) \times 10^{16} \ \rm G$. 
We fix other parameters as $B_{\rm dip} = 2 \times 10^{13} \ \rm G$, $P_{\rm i} = 1 \ \rm ms$, and $M_{\rm ej} = 2 \ M_\odot$. 
}
\label{fig:x_lc_gw}
\end{figure}

\begin{figure}
\centering
\includegraphics[width=90mm]{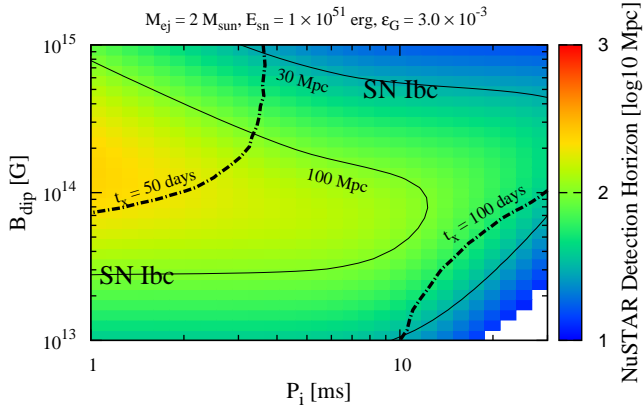}
\caption{
Contour plot showing properties of hard X-ray counterparts of fast-spinning strongly magnetized proto-NS formation 
with $M_{\rm ej} = 2 \ M_\odot$, $M_{\rm {}^{56}Ni} = 0.05 \ M_\odot$, ${\cal E}_{\rm sn} = 1 \times 10^{51} \ \rm erg$, and $\epsilon_{\rm G} = 3.0 \times 10^{-3}$. 
The color with solid lines corresponds to the detection horizon using {\it NuSTAR} with $50 \ \rm ks$ observations at around the peak time shown in the dotted-dashed lines. 
}
\label{fig:x_mej2_bt3e16}
\end{figure}

Fig. \ref{fig:x_lc_gw} shows the hard X-ray counterpart of the SN shown in Fig. \ref{fig:sn_height_gw}. 
The hard X-ray counterpart is also more suppressed for a larger GW spin-down.   
%However, we should note that, comparing the pulsar-driven models compatible with SN Ibc in Fig. \ref{fig:x_lc} and Fig. \ref{fig:x_lc_gw}, the latter case give a larger hard-X-ray flux. 
%This is essentially because that the latter case has a longer dipole spin-down time, and can give a larger electromagnetic luminosity in later phases. 
A broader parameter space is investigated in Fig. \ref{fig:x_mej2_bt3e16} for $M_{\rm ej} = 2 \ M_\odot$ cases. 
The hard X-ray counterpart of  pulsar-driven SN Ibc with strong GW spin-down can be detectable even from $\sim 100 \ \rm Mpc$ by follow-up observations $\sim 50-100 \ \rm days$ after the explosion using {\it NuSTAR}. 
The maximum detectable event rate both by the second generation GW detector network and {\it NuSTAR} is $\sim 1 \ \rm yr^{-1} \ sky^{-1}$. 

\subsection{Neutrino emission}
We briefly overview possible neutrino counterparts from fast-spinning newborn NSs. 
Neutrinos can be used as a powerful probe of fast-rotating pulsars, because they can escape without significant attenuation.  
Multi-MeV thermal neutrinos from a nearby SN are detectable with Hyper-Kamiokande~\citep{HK_2011}. 
In addition, if the proto-NS is rotating and magnetized, nucleons supplied by the neutrino-driven wind should be accelerated magnetically, 
leading to quasi-thermal neutrino emission in the GeV-TeV range~\citep{Murase_et_al_2014}.  
Even higher-energy neutrinos can be produced depending on the magnetic dissipation process~\citep{Murase_et_al_2009,Fang_et_al_2014,Lemoine_et_al_2015}. 
If the pair-multiplicity is not as much large, PeV-EeV neutrinos, which can be produced by $pp$ and/or $p\gamma$ interactions, can be detected up to $\sim10$~Mpc.  
Note that high-energy neutrino emission is typically expected for SNe II 
when the SN shock becomes collisionless after its shock breakout~\citep{Katz_et_al_2011,Murase_et_al_2011,Kashiyama_et_al_2013}.
%The situation can be different in pulsar-driven SNe, where particle acceleration may work even before the shock breakout and without dense circumstellar material or extended envelope.  
%Thus, detections of high-energy neutrinos from Type Ibc SNe would suggest that ions are accelerated in the pulsar wind or choked jet inside the SN ejecta. 

\section{Discussion}\label{sec:discussion}

\subsection{Observational strategies}
Based on the results in the previous sections, we discuss possible observational strategies for fast-spinning strongly magnetized newborn NSs in SE SNe. 

\subsubsection{SNe Ibc} 
The optical light curve of SN Ibc can be broadly consistent with the pulsar-driven SN model with $P_{\rm i} \gtrsim 10 \ \rm ms$, $B_{\rm dip} \gtrsim 5 \times 10^{14} \ \rm G$ and  $M_{\rm ej} \lesssim 5 \ M_\odot$. 
For such cases, {\it NuSTAR} can detect the hard X-ray PWN emission from $d_{\rm L} \lesssim 30 \ \rm Mpc$.  
We encourage follow-up observations of $\sim 50 \ \rm ks$, $\sim 50-100 \ \rm days$ after the SN explosion. 
The anticipated SN Ibc rate within the {\it NuSTAR} detection horizon is about $\rm a \ few \ yr^{-1}$ (see Fig. \ref{fig:rate}). 

If the inner toroidal magnetic field is as strong as $B_{\rm t} \gtrsim 10^{16} \ \rm G$, the GW spin-down due to magnetically deformed rotation can effectively suppress the pulsar wind, 
and the pulsar-driven SN emission can be consistent with SNe Ibc if $P_{\rm i}$ is about $\rm a \ few \ ms$ and $B_{\rm dip} \lesssim 10^{14} \ \rm G$. 
Such GWs from SNe Ibc can be detectable up to $\lesssim 20 \ \rm Mpc$ by Advanced LIGO. 
{\it NuSTAR} can also detect the hard X-ray counterpart with a same follow-up observation as above. 
Note that the typical GW emission duration is $\sim 0.1 \ \rm days$,  
which is much shorter than the relevant timescales of the SN and hard X-ray counterpart, $\gtrsim 10 \ \rm days$.
In order to pin down the GW time window, it will be useful to combine neutrino emission~\citep[e.g.,][]{Murase_et_al_2009,Murase_et_al_2014} 
or some electromagnetic precursor signals~\citep[e.g.,][]{Kistler_et_al_2013,Nakauchi_et_al_2015}.   

Multi-messenger detections or even non-detections from nearby SNe Ibc give important implications on the origin of Galactic magnetars.  
If the progenitors of Galactic magnetars are WRs and the formation rate is as high as Eq. (\ref{eq:magnetar}), 
ordinary SN Ibc is the promising site of the magnetar formation. 
Detections of the hard X-ray counterpart can fix where and how Galactic magnetars are formed. 
On the other hand, non-detection may indicate a lower formation of magnetar rate than Eq. (\ref{eq:magnetar}) 
or different formation pathway, e.g., the ``fossil" scenario~\citep{Usov_1992,Thompson_Duncan_1993,Ferrario_Wickramasinghe_2006}.  
%Also note that relatively slow rotation periods of $P_{\rm i} \sim 1 \ \rm s$ are inferred for several galactic magnetars with supernova remnants 
%based on the conventional dipole spin-down formula~\citep{Suwa_Enoto_2014}.
Moreover, the multi-messenger detections can strongly constrain the initial spin of proto-NS, which is a missing link in the massive star evolution; 
spins of massive O-type stars and NS pulsars have been measured~\citep{Fukuda_1982}, but no observational information about the in-between. 
Probing the proto-NS spin will shed light on the angular-momentum-transfer process during massive star evolution, which is still highly uncertain~\citep[e.g.,][]{Meynet_et_al_2011}.  

\subsubsection{BL-SNe Ibc} 
The optical light curve of BL-SN Ibc can be broadly consistent with the pulsar-driven SN model with $P_{\rm i}$ less than $\rm a \ few \ ms$, $B_{\rm dip} \gtrsim 5 \times \ 10^{14} \ \rm G$, and  $M_{\rm ej} \gtrsim 5 \ M_\odot$. 
For such cases, {\it NuSTAR} can detect the hard X-ray PWN emission from $d_{\rm L} \lesssim 50 \ \rm Mpc$.  
It requires a follow-up observation of $\sim 50 \ \rm ks$, $\sim 50-100 \ \rm days$ after the SN explosion. 
The BL-SN Ibc rate within the {\it NuSTAR} detection horizon is $\sim 1 \ \rm yr^{-1}$ (see Fig. \ref{fig:rate}). 

Detections of such hard X-ray counterparts will indicate that newborn pulsars play relevant roles in BL-SNe Ibc.
Spin-down power of newborn pulsars can be as important as ${}^{56}$Ni decay for the SN emission. 
Also, the large kinetic energy inferred for BL-SN Ibc ejecta can be mainly provided by the spin-down power. 
The pulsar-driven SN model for BL-SN Ibc requires a magnetar-class field strength and $P_{\rm i} \sim \rm ms$.  
So, the hard X-ray observation can probe the nascent stage of magnetars, namely indicating that the ms rotation is relevant for the magnetic field amplification.  
We should note, however, that a Galactic magnetar remnant is known to be less energetic, ${\cal E}_{\rm K} \lesssim 10^{51} \ \rm erg$~\citep{Vink_Kuiper_2006}. 
Also, the observed BL-SN Ibc rate is smaller than the magnetar formation rate, thus BL-SNe Ibc could explain only a minor abundance of Galactic magnetars.

\subsubsection{SL-SNe Ic} 
The pulsar-driven SN model can explain the optical light curve of SL-SN Ic with $P_{\rm i}$ less than $\rm a \ few \ ms$ and $B_{\rm dip} \gtrsim 10^{13} \ \rm G$. 
%The ejecta mass is relatively large, $M_{\rm ej} \gtrsim 5 \ M_\odot$.  
For such cases, the hard X-ray counterpart can be detectable from $d_{\rm L} \lesssim \rm Gpc$ ($z \lesssim 0.2$) using {\it NuSTAR} with $\sim 50 \ \rm ks$ observations. 
A follow-up observation $\sim 100 \ \rm day$ after the SN explosion is required. 
Current optical transient surveys like PTF~\citep{Law_et_al_2009}, Pan-STARRs~\citep{Hodapp_et_al_2004}, and ASASSN~\citep{Shappee_et_al_2014} 
can detect the optical counterpart with a rate larger than $\rm a \ few \  yr^{-1}$ from within the {\it NuSTAR} detection horizon. 

Detections of the hard X-ray counterpart can strongly support the pulsar-driven SN model for SL-SNe Ic. 
We should note that early PWN emissions could be also observed in soft X rays~\citep{Metzger_et_al_2013}, GeV-TeV gamma rays~\citep{Kotera_et_al_2013,Murase_et_al_2015}, and radio, 
but the detectability will be more sensitive depends on, e.g., the ionization fraction of the ejecta and the lepton acceleration in the PWN.

\subsection{Impacts of Simplifications}\label{sec:caveats}
Our semi-analytic model includes several simplifications, which needs to be refined for more detailed comparison with observation. 
The SN light curve is calculated based on the one-zone approximation.  
Including the multi-dimensional effects is crucial to obtain multi-band light curve more precisely.  
We effectively fix the opacity of the elastic scattering and photoelectric absorption separately, but these quantities need to be determined consistently in a time-dependent manner. 
More detailed radiation-transfer calculations taking into account the ionization degree of metals are required.  
Our treatment basically overestimates the photoelectric absorption, and thus the SN counterpart, and underestimates the soft X-ray counterpart, although the effect is minor for the SN counterpart. 
As for the injection spectrum from PWNe, we use a simple broken power law motivated by detailed numerical calculations by \cite{Murase_et_al_2015}. 
However, the shape of the spectrum in general changes with time depending on the Compton parameter $Y$ in the nebula. 
Our treatment overestimates the gamma-ray flux once $Y$ becomes small.

\section{Summary}
To test the pulsar-driven supernova~(SN) scenario for stripped-envelope~(SE) SNe from broad-line SNe Ibc to super-luminouse SNe Ic and ordinary SN Ibc, and also the Galactic magnetar connection to SE-SNe, 
we calculate multi-messenger counterpart of fast-spinning strongly magnetized proto-neutron star~(NS) formation in massive collapses with SE.  
The SN emission powered by pulsar spin-down and ${}^{56}$Ni-decay, early pulsar-wind-nebular~(PWN) emission, and gravitational wave~(GW) spin-down are consistently modeled. 

We show that  the peak light curves of all types of observed SE-SNe, can be broadly explained by the pulsar-driven SN model;  
$P_{\rm i} \gtrsim 10 \ \rm ms$ and $B_{\rm dip} \gtrsim 5 \times \ 10^{14} \ \rm G$ for SN Ibc, 
$P_{\rm i}$ less than $\rm a \ few \ ms$ and $B_{\rm dip} \gtrsim 5 \times \ 10^{14} \ \rm G$ for BL-SN Ibc, 
and $P_{\rm i}$ less than $\rm a \ few \ ms$ and $B_{\rm dip} \gtrsim 10^{13} \ \rm G$ for SL-SN Ic. 
The latter two cases prefer more massive progenitors. 

For all cases, the early PWN emission especially in the hard X-ray band can be the smoking-gun signal of an underlying newborn pulsar engine, 
detectable by follow-up observations using {\it NuSTAR} $\sim 50-100 \ \rm days$ after the explosion. 
The hard X-ray detection horizon is $\lesssim 30 \ \rm Mpc$ for SN Ibc and BL-SN Ibc, and $\lesssim 1 \ \rm Gpc$ for SL-SN Ic, and the potential detection rates are $\sim 1 \ \rm yr^{-1} \ sky^{-1}$. 

If the inner toroidal magnetic field is as strong as $B_{\rm t} \gtrsim 10^{16} \ \rm G$, the GW spin-down due to magnetically deformed rotation can be relevant 
especially for the cases with $P_{\rm i}$ about $\rm a \ few \ ms$ and $B_{\rm dip} \lesssim 10^{14} \ \rm G$. 
The GW counterpart can be detectable up to $\lesssim 20 \ \rm Mpc$ by the second generation GW detector network. 
When the GW spin-down is strong enough to be detected, the pulsar-driven SN cannot be as bright as SL-SN Ic; 
instead, millisecond proto-NSs with $B_{\rm dip} \sim {\rm a \ few} \times 10^{13} \ \rm G$ result in ordinary SNe Ibc.

\acknowledgments
When this work was completed, we became aware of the related, independent work, arXiv:1508.02712.  
While a part of the interest is shared, our work has more focus on the scenarios explaining both broad-lined supernovae and superluminous supernovae, and multi-messenger prospects.  

This work is supported by NASA through Einstein Postdoctoral Fellowship grant number PF4-150123 awarded by the Chandra X-ray Center, 
which is operated by the Smithsonian Astrophysical Observatory for NASA under contract NAS8-03060 (K. Kashiyama).
K. M. acknowledges continuous support from Institute for Advanced Study and Pennsylvania State University. 
IB is thankful for the generous support of Columbia University in the City of New York and the National Science Foundation under cooperative agreement PHY-1404462.
K. Kiuchi is supported by a Grant-in-Aid for Scientific Research (15H00783,15H00836,15K05077) and by HPCI Strategic Program of Japanese MEXT/JSPS (Project numbers hpci130025, 140211, and 150225).

\appendix
In this Appendix, we show the semi-analytic model we use for calculating the multi-messenger counterpart of fast-spinning strongly magnetized NS formation in massive star core collapse.  

\section{Spin-down}\label{sec:sd}
The spin-down of proto-NS is calculated from~\citep{Ostriker_Gunn_1969}    
\begin{equation}\label{eq:L}
-\frac{d{\cal E}_{\rm rot}}{dt} = L_{\rm em} + L_{\rm gw}.
\end{equation}
The first term in the right-hand side is the magnetic spin-down power; 
\begin{equation}\label{eq:L_m_1}
L_{\rm em} = \frac{\mu^2 (2\pi/P)^4}{c^3} (1+ C \sin^2 \chi_\mu), 
\end{equation}
%\begin{equation}
%\mu = \frac{1}{2} B_{\rm dip} R^3 \sim 8.6 \times 10^{31} \ B_{\rm p, 14} \ \rm G \ cm^3,
%\end{equation}
where $\mu = B_{\rm dip} R^3/2$ is the magnetic moment, $P$ is the rotation period, $\chi_\mu$ is the angle between the magnetic and rotation axis, and $C \sim 1$ is a pre-factor. 
Eq. (\ref{eq:L_m_1}) is obtained by force-free simulations~\citep{Gruzinov_2005,Spitkovsky_2006,Tchekohovskoy_et_al_2013}.  
Note that Eq. (\ref{eq:L_m_1}) is larger than the classical dipole spin-down luminosity by $3(1+C \sin^2 \chi_\mu)/2\sin^2\chi_\mu > 1$. 
We assume that the magnetized wind is isotropic for simplicity.  
These approximations are not valid within the Kelvin-Helmholtz timescale $t_{\rm KH, \nu} \lesssim 100 \ \rm s$  
where the baryon loading on the magnetized wind via neutrino-driven wind from the proto-NS surface is relevant~\citep[e.g.,][]{Thompson_et_al_2004}.  
Also, for an extremely strong poloidal field, $B_{\rm dip} \gtrsim 10^{15} \ \rm G$, ms-proto-NSs spin down within the KH timescale. 
In such cases, the magnetized wind could punch out the progenitor star as a bi-polar jet collimated by the anisotropic stress and the hoop stress~\citep{Buccianitini_et_al_2007,Buccianitini_et_al_2008}. 
We here only consider a longer timescale $t \gg t_{\rm KH, \nu}$ and a poloidal field $B_{\rm dip} < 10^{15} \ \rm G$. 

The second term in the right-hand side of Eq. (\ref{eq:L}) represents the GW energy loss; 
\begin{equation}\label{eq:L_gw_1}
L_{\rm gw} = \frac{2}{5}\frac{G(\epsilon_{\rm G} I)^2 (2 \pi/P')^6}{c^5} \sin^2 \chi_{\epsilon_{\rm G}}(1+15 \sin^2 \chi_{\epsilon_{\rm G}}),  
\end{equation}
where $\epsilon_{\rm G} \equiv \Delta I/I$ is the deformation rate, $P'$ is the pattern period, and $\chi_{\epsilon_{\rm G}}$ is the angle between the deformation and rotation axis~\citep{Cutler_Jones_2001}.  
%Fast-rotating proto-NSs can be also deformed by their own rotation. 
%However, this contribution is effectively spherically symmetric, and thus irrelevant for the GW emission~\citep{Jones_Andersson_2001}.  
In this paper, we consider the magnetically deformed rotation~\citep{Cutler_2002,Stella_et_al_2005,DallOsso_et_al_2009}, 
which is an interesting channel especially in terms of magnetar formation. 
Once inner toroidal magnetic fields amplified up to a magnetar value, the proto-NS becomes oblate by the magnetic pinch~\citep[see, e.g.,][]{Cutler_2002,Kiuchi_Yoshida_2008,Gualtieri_et_al_2011}. 
The deformation rate can be estimated as 
\begin{equation}\label{eq:epsilon}
\epsilon_{\rm G} = \frac{15}{4} \frac{{\cal E}_{\rm B}}{|W|}  \sim 2.5 \times 10^{-4} \left(\frac{B_{\rm t}}{10^{16} \ \rm G}\right)^2. 
\end{equation}  
Here, $|W| \approx  M_{\rm ns} c^2 \times 0.6 {\cal C}/(1-0.5{\cal C}) \sim 4.4 \times 10^{53} \ \rm erg$ is the gravitational energy of the proto-NS 
with compactness parameter, ${\cal C} = GM_{\rm ns}/R_{\rm ns}c^2 \sim 0.17$~\citep{Lattimer_Prakash_2001}. 
%If the equipartition is realized between the magnetic and rotation energies, the inner toroidal magnetic field can be as strong as 
%$(B_{\rm t, max}^2/8\pi) \times (4\pi R_{\rm ns}{}^3/3) \approx {\cal E}_{\rm rot, i}$, or 
%\begin{equation}
%B_{\rm t, max} \sim 3.1 \times 10^{17} \ P_{0, -3}{}^{-1} \ \rm G. 
%\end{equation}
In general, the deformation axis is not coincide with the rotation axis, and the proto-NS precesses around the rotation axis~\citep{Mestel_Takhar_1972}. 
%with a timescale of $t_{\rm pre} \approx P_{\rm i}/\epsilon \cos \chi_0$ if the system is non-dissipative and adiabatic. 
The tilt angle of the precession can increase secularly due to the bulk viscosity~\citep{DallOsso_et_al_2009}, %\footnote{
%Note that the GW radiation reaction reduces the tilt angle, and the proto-magnetar can become oblate ($\chi \approx 0$) 
%with a timescale of $t_{rr} \sim 45 \ B_{t, 16}{}^{-4} P_{0, -3}{}^4 \ \rm days$~\citep{Cutler_Jones_2001}. 
%The effect of the GW radiation reaction can be neglected if e.g., $t_{vis} < t_{rr}/10$, or $B_{t} < 9 \times 10^{16} \ {\rm G} \ T_{pns, 10} P_{0, -3}{}^{1/3}$
%for $\chi_0 \approx 0$ 
and the proto-NS evolves into a prolate shape, which is a plausible configuration for the GW emission ($\chi_\mu = \chi_{\epsilon_{\rm G}} = \pi/2$, $P = P'$). 
%\begin{equation}\label{t_vis}
%t_{vis} \sim 3.9 \ (1+3\cos^2\chi_0)^{-1} B_{t, 16}{}^2 P_{0, -3}{}^2 T_{pns, 10}{}^{-6} \ \rm s.  
%\end{equation}
%Here, $T_{pns} = 10^{10} \ T_{pns, 10} \ \rm K$ is the temperature of the proto-magnetar, which can also time-evolve, and need to be solved consistently.    
%Note also that the above expression for $t_{vis}$ is obtained by assuming that the angular momentum of the proto-magnetar is constant. 
%The magnetic spin-down barely affect the dumping of the precession if, e.g., $t_{vis} < t_{sd, m}/10$ following \cite{DallOsso_et_al_2009}, which can be expressed as 
%\begin{equation}\label{eq:dump}
%B_{t} < 2.4 \times 10^{16} \ {\rm G} \ P_{0, -3}{}^{-1} [\ln(320 \times P_{0, -3}{}^2 B_{p, 14}{}^{-2}+1)]^{1/2}.
%\end{equation}
%with an appropriate assumption on the time-evolution of the proto-NS temperature~\citep{DallOsso_et_al_2009}. 
Recently, a precessing motion driven by deformation as Eq. (\ref{eq:epsilon}) was inferred for a galactic magnetar from the X-ray timing observation~\citep{Makishima_et_al_2014, Makishima_et_al_2015}. %\footnote{
%\bf The maximum strain that a NS crust can support is estimated to be $\epsilon \lesssim 10^{-5}$~\citep[e.g.,][]{Shaham_1977}. 
%What does this mean? Maybe, the crust of 4U 0142+61 is also squeezed.}
This GW emission can only occur if the viscous dumping timescale of the NS precession is shorter than the competitive magnetic braking timescale. 
This condition can be described as~\citep{DallOsso_et_al_2009}
\begin{equation}\label{eq:dump}
B_{\rm t} < 2.4 \times 10^{16} \ {\rm G} \ \left(\frac{P_{\rm i}}{\rm ms}\right)^{-1} \left( \ln \left[320 \left(\frac{P_{\rm i}}{\rm ms}\right)^2 \left(\frac{B_{\rm dip}}{10^{14} \ \rm G}\right)^{-2}+1\right]\right)^{1/2}.
\end{equation} 

\section{Dynamics}
We here describe a simple model for dynamics of the SN ejecta and the resulting electromagnetic emission. The radius of the ejecta evolves as
\begin{equation}\label{eq:R}
\frac{dR_{\rm ej}}{dt} = V_{\rm ej}.
\end{equation} 
We assume the density structure of the SN ejecta as 
\begin{equation}
\rho_{\rm ej} \approx \frac{3-\delta}{4 \pi} \frac{M_{\rm ej}}{R_{\rm ej}^3} \left(\frac{R}{R_{\rm ej}}\right)^{-\delta},
\end{equation} 
We take $\delta = 1$ as a fiducial value for the index, and thus a dominate fraction of mass resides at around $R \approx R_{\rm ej}$.
%We note that the initial radius of the ejecta, $R_{\rm ej, 0} \approx R_\ast$ ($\sim 10^{11} \ \rm cm$ for WRs), $t > R_{\rm ej, 0} /v_{\rm ej, 0} \sim 100 \ E_{\rm sn, 51}{}^{-1/2} M_{\rm ej, 0}{}^{1/2} R_{\rm ej,0,11} \ \rm s$. 
Without significant energy injection after the explosion, the expansion velocity is almost constant, 
$\approx V_{\rm ej, i} = (2{\cal E}_{\rm sn}/M_{\rm ej})^{1/2} \sim 10,000 \ {\rm km \ s^{-1}} \ ({\cal E}_{\rm sn}/10^{51} \ {\rm erg})^{1/2} (M_{\rm ej}/M_\odot)^{-1/2} \ \rm cm \ s^{-1}$ up to the Sedov radius.
On the other hand, when a fast-spinning proto-NS exists, the ejecta is accelerated by the magnetized wind, 
\begin{equation}\label{eq:E_kin}
\frac{d{\cal E}_{\rm K}}{dt} = \frac{{\cal E}_{\rm int}}{t_{\rm dyn}},
\end{equation}
where ${\cal E}_{\rm K} \approx M_{\rm ej} V_{\rm ej}/2$ is the kinetic energy, ${\cal E}_{\rm int}$ is the total internal energy, and 
\begin{equation}
t_{\rm dyn} = \frac{R_{\rm ej}}{V_{\rm ej}} 
\end{equation}
is the dynamical timescale of the ejecta. 
The energy injection from the underlying pulsar occurs at the shock between pulsar wind and the SN ejecta. 
The radius of the shocked wind region evolves as 
\begin{equation}
\frac{dR_{\rm w}}{dt} = V_{\rm nb} + \frac{R_{\rm w}}{t}.
\end{equation}
Here $V_{\rm nb}$ is the expansion velocity, which is obtained from the pressure equilibrium, 
$\int L_{\rm em} \times  \min[1, \tau_{\rm T}^{\rm ej} V_{\rm ej}/c]  dt/ 4\pi R_{\rm w}^3 \approx 6 \rho_{\rm ej} V_{\rm nb}^2/7$, or
\begin{equation}
V_{\rm nb} \approx \sqrt{\frac{7}{6(3 -\delta)} \frac{\int L_{\rm em} \times  \min[1, \tau_{\rm T}^{\rm ej} V_{\rm ej}/c]  dt}{M_{\rm ej}}\left(\frac{R_{\rm ej}}{R_{\rm w}}\right)^{3-\delta}}.
\end{equation}
The factor $\min[1, \tau_{\rm T}^{\rm ej} V_{\rm ej}/c]$ represents the fraction of the spin-down luminosity deposited in the SN ejecta (see Eq. \ref{eq:tau_es} for the definition of $\tau_{\rm T}^{\rm ej}$). 
If $R_{\rm w}  \geq R_{\rm ej}$, we set $R_{\rm w} \approx R_{\rm ej}$.

\section{Electromagnetic Emission}\label{sec:em}
The time evolution of the internal energy in the SN ejecta is described as 
\begin{equation}\label{eq:Eint}
\frac{d{\cal E}_{\rm int}}{dt} = - L_{\rm sn} - \frac{{\cal E}_{\rm int}}{t_{\rm dyn}} + f_{\rm dep, em} L_{\rm em} + f_{\rm dep, {}^{56}Ni} L_{\rm {}^{56}Ni}+ f_{\rm dep, {}^{56}Co} L_{\rm {}^{56}Co}.
\end{equation}
The first and second terms on the right-hand side correspond to the energy loss via quasi-thermal SN emission and adiabatic expansion, 
whereas the third, foruth, and fifth terms correspond to the energy injection via magnetar wind, ${}^{56}$Ni and Co decay, respectively. 

The bolometric SN luminosity can be given by
\begin{equation}
L_{\rm sn} \approx \frac{{\cal E}_{\rm int}}{t_{\rm esc}^{\rm ej}}, 
\end{equation}
where
\begin{equation}
t_{\rm esc}^{\rm ej} = \tau_{\rm T}^{\rm ej} \frac{R_{\rm ej}}{c},   
\end{equation}
is the thermal photon escape time from the ejecta, 
\begin{equation}\label{eq:tau_es}
\tau_{\rm T}^{\rm ej} = \frac{(3-\delta)K_{\rm T} M_{\rm ej}}{4 \pi R_{\rm ej}{}^2},
\end{equation}
is the optical depth of the ejecta, $K_{\rm T} = \xi \sigma_{\rm T}/\mu_{\rm e} m_u$ is the Thomson opacity, 
$\mu_{\rm e}$ is the mean molecular weight per electron, $m_u = 1.66 \times 10^{-24} \ \rm g$ is the atomic mass unit, and $0 \leq \xi \leq 1$ is the effective ionization fraction. 
Since we are mainly interested in SE SNe, we take $\mu_{\rm e} = 2$. 
In general, $\xi$ depends on the temperature and composition and evolves with time. 
Here, for simplicity, we use fixed values $\xi = 0.25-1$, i.e.,  $K_{\rm T} \sim 0.05 - 0.2 \ \rm g^{-1} \ cm^2$, 
which is reasonable at around the optical peak of SE SNe.  
The temperature of the emission can be estimated as
\begin{equation}
T_{\rm sn} = \left(\frac{{\cal E}_{\rm int}}{a{\cal V}_{\rm ej}}\right)^{1/4}
\end{equation}
where ${\cal V}_{\rm ej} \approx 4 \pi R_{\rm ej}^3/3$ and $a$ is the radiation constant. 
Note that the above method of calculating the SN emission is equivalent to the classical Arnett model~\citep{Arnett_1982} with uniform ejecta temperature~\citep{Chatzopoulos_et_al_2012}.

At the interface of the magnetized wind and SN ejecta, highly relativistic electrons are injected, 
which are further accelerated, e.g., at the shock or in the magnetic turbulence, and then rapidly cool via synchrotron emission and inverse Compton scattering.
The scattered photons have very high energies so that they can produce pairs by two-photon annihilation.
The synthesized electron/positron is still energetic and produces another pair successively.     
\cite{Murase_et_al_2015} numerically calculated the above electromagnetic cascade process by assuming the electron injection spectrum as 
 \[ 
  \frac{d\dot{N}_{\rm e}^{\rm inj}}{d \gamma_{\rm e}} \propto \begin{cases}
      (\gamma_{\rm e}/ \gamma_{\rm b})^{-q_1} & (\gamma_{\rm e} < \gamma_{e, \rm b}),  \\
      (\gamma_{\rm e}/ \gamma_{\rm b})^{-q_2} & (\gamma_{\rm b} < \gamma_{\rm e} < \gamma_{\rm M}),
  \end{cases}
\]
with $q_1 \sim 1-1.5$, $q_2 \sim 2.5-3$, $\gamma_{\rm b} \sim 10^{4.5-6}$, which is motivated by the observation of young PWNe~\citep{Tanaka_Takahara_2010}. 
The electron maximum energy can be estimated by equating the acceleration timescale $t_{\rm acc} = \eta \gamma_{\rm e} m_e c^2/eBc$ 
and synchrotron cooling timescale $t_{\rm syn} = 3m_{\rm e}c/4 \sigma_{\rm T} U_B \gamma_{\rm e}$, i.e., $\gamma_{\rm M} \approx (6 \pi e/\eta \sigma_T B)^{1/2}$. 
Here, the magnetic field energy density is given by  
\begin{equation}
U_B = \frac{B^2}{8 \pi} = \epsilon_B \frac{3 \int L_{\rm em} dt}{4 \pi R_{\rm w}^3},
\end{equation}
with $\epsilon_B \sim 10^{-(2-3)}$. 
\cite{Murase_et_al_2015} found that the resultant nebula spectrum can be well approximated by a broken power law; 
\[
  E_\gamma \frac{dN_\gamma}{d E_\gamma} \approx \frac{\epsilon_{\rm e} L_{\rm em}}{{\cal R}_{\rm b} E_{\rm syn}^{\rm b}} \begin{cases}
      (E_\gamma/ E_{\rm syn}^{\rm b})^{-q_1/2} & (E_\gamma < E_{\rm syn}^{\rm b}),  \\
      (E_\gamma/ E_{\rm syn}^{\rm b})^{-1} & (E_{\rm syn}^{\rm b}< E_\gamma < \varepsilon_{\gamma, \rm max}), 
  \end{cases}
\]  
where ${\cal R}_{\rm b} \sim 2/(2-q_1) + \ln(\varepsilon_{\gamma, \rm max}/E_{\rm syn}^{\rm b})$ and $\epsilon_{\rm e} = 1 - \epsilon_B \approx 1$. 
The low-energy part is dominated by the fast-cooling synchrotron emission from injection electrons with $\gamma_{\rm e} < \gamma_{\rm b}$.  
The corresponding break photon energy is 
\begin{equation}
E_{\rm syn}^{\rm b} = \frac{3}{2} \hbar \gamma_{\rm b}^2 \frac{eB}{m_{\rm e} c}. 
\end{equation}
On the other hand, the high-energy part is mainly produced by inverse Compton scattering and successive pair cascade. 
The maximum energy is determined by the two-photon annihilation with the SN photons, 
\begin{equation}
E_\gamma^{\rm M} \approx \frac{m_{\rm e} c^2}{2k_{\rm B} T_{\rm sn}}. 
\end{equation}

Injected non-thermal photons from the wind nebula are down-scattered, or absorbed during propagating through the SN ejecta.
The main interaction channel depends on the photon energy; 
Bethe-Heitler~(BH) pair production for $E_\gamma \gtrsim 10 \ \rm MeV$,
Compton scattering for $10 \ {\rm keV} \lesssim E_\gamma \lesssim \rm 10 MeV$,
photoelectric (bound-free) absorption for $10 \ {\rm eV} \ \lesssim E_\gamma \lesssim 10 \ \rm keV$, 
bound-bound and free-free absorption for lower energy bands. 
We calculate the energy deposition fraction of a photon as 
\begin{equation}\label{eq:f_dep}
f_{\rm dep} = \max[1, f_{\rm dep, sc}+f_{\rm dep, ab}]. 
\end{equation}
The contribution from the Compton scattering is estimated as 
\begin{equation}
f_{\rm dep, sc} = 1 - (1-K_{\rm comp})^{\max[\tau_{\rm comp}, \tau_{\rm comp}^2]}, 
\end{equation}
where $K_{\rm comp}$ is the inelasticity of Compton scattering,
%\begin{eqnarray}
%K_{\rm comp} \sigma_{\rm comp} &=& \frac{3}{4} \sigma_{\rm T} \left[ \frac{2(1+x)^2}{x^2(1+2x)} -\frac{1+3x}{(1+2x)^2} - \frac{(1+x)(2x^2-2x-1)}{x^2(1+2x)^2} - \frac{4x^2}{3(1+2x)^3} \right. \\ \notag
%&& \left. -\left(\frac{1+x}{x}-\frac{1}{2x} +\frac{1}{2x^3}\right) \ln (1+2x) \right].
%\end{eqnarray}
$\sigma_{\rm comp}$ is the Klein-Nishina cross section, 
%\begin{eqnarray}
%\sigma_{\rm comp}  &=& \frac{3}{4} \sigma_{\rm T} \left[\frac{1+x}{x^3}\left\{\frac{2x(1+x)}{1+2x}-\ln(1+2x)\right\}+\frac{\ln(1+2x)}{2x}-\frac{1+3x}{(1+2x)^2}\right],
%\end{eqnarray} 
and 
%\begin{equation}
%\tau_{\rm comp} = \frac{(3-\delta)\sigma_{\rm comp} M_{\rm ej}}{4 \pi \mu_e m_u R_{\rm ej}{}^2}.
%\end{equation}
$\tau_{\rm comp}$ is the optical depth~\citep[see][]{Murase_et_al_2015}. %with $x = E_\gamma/m_e c^2$. 
On the other hand, the energy deposition fraction by the absorption processes can be expressed as
\begin{equation}
f_{\rm dep, ab} = 1 - \exp(-\tau_{\rm ab}),
\end{equation}
where 
\begin{equation}
\tau_{\rm ab} = \tau_{\rm BH} + \tau_{\rm pe}, 
\end{equation}
and $\tau_{\rm BH}$ and $\tau_{\rm pe}$ are the optical depth of BH pair production and photoelectric absorption. 
We treat the BH pair production as an absorption process since the inelasticity is $K_{\rm BH} = 1 - 2/(E_\gamma/m_{\rm e}c^2) \approx 1$ for $E_\gamma/m_{\rm e}c^2 \gg 2$~\citep[see][]{Murase_et_al_2015}.
%The optical depth can be estimated as 
%\begin{equation}
%\tau_{\rm BH} \approx \frac{(3-\delta)(Z_{\rm eff} +1)\sigma_{\rm BH, p} M_{\rm ej}}{8\pi m_u R_{\rm ej}{}^2},
%\end{equation}
%with the cross section being approximately given by 
%\begin{equation}
%\sigma_{\rm BH, p} \approx \frac{3 \alpha}{8\pi} \sigma_{\rm T} \left[\frac{28}{9}\ln(2x) + \frac{218}{27}\right].
%\end{equation}
%Note that e.g., $Z_{\rm eff} \approx 7$ for $X_{\rm CO} \approx 1$. 
The optical depth of the photoelectric absorption can be estimated as  
\begin{equation}
\tau_{\rm pe} \approx \frac{(3-\delta) K_{\rm pe}M_{\rm ej}}{4 \pi R_{\rm ej}^2},
\end{equation} 
where we use an approximate form of the opacity for oxygen-dominated ejecta, 
\begin{equation}
K_{\rm pe} \approx 5.0 \times \zeta \left(\frac{E_\gamma}{10 \ \rm keV}\right)^{-3} \ \rm g^{-1} \ cm^2, 
\end{equation}
Here $0 \leq \zeta \leq 1$ is a time-dependent factor determined from the effective ionization fraction.  
In this paper, we fix $\zeta = 0.5$ for simplicity.  
Note that the fraction of the energy in the soft X-ray band (and lower energy bands) is always subdominant in our case.
For example, the SN light curve around the peak does not change significantly depending on the details of the photoelectric absorption. 
The photoelectric absorption can be relevant for the SN light curve in the late phase ($\gtrsim 100 \ \rm days$ after the explosion) and the soft X-ray light curve. 

The total energy deposition fraction of the magnetized wind is calculated by 
\begin{equation}
f_{\rm dep, em} = \frac{\int  f_{\rm dep}(E_\gamma) E_\gamma \frac{dN_\gamma}{dE_\gamma} dE_\gamma}{\int E_\gamma \frac{dN_\gamma}{dE_\gamma} dE_\gamma}, 
\end{equation}
where $dN/dE_\gamma$ is the wind nebula spectrum and $f_{\rm dep} (E_\gamma)$ is the energy deposition fraction of a photon with an energy $E_\gamma$. 
On the other hand, the observed non-thermal nebula spectrum can be calculated as 
\begin{equation}
 E_\gamma \frac{dN_{\gamma \rm obs}}{d E_\gamma} \approx f_{\rm esc} \times E_\gamma \frac{dN_\gamma}{d E_\gamma}, 
\end{equation}
where 
\begin{equation}\label{eq:f_esc}
f_{\rm esc} = \prod_{\rm X = comp, ab} \{\exp(-\tau_{\rm X}) + [1-\exp(-\tau_{\rm X})](1-K_{\rm X})^{\max[\tau_{\rm X},\tau_{\rm X}^2]}\}.
\end{equation}
is the fraction of injected PWN emission directly escape from the SN ejecta. 

\begin{figure}
\centering
\includegraphics[width=90mm]{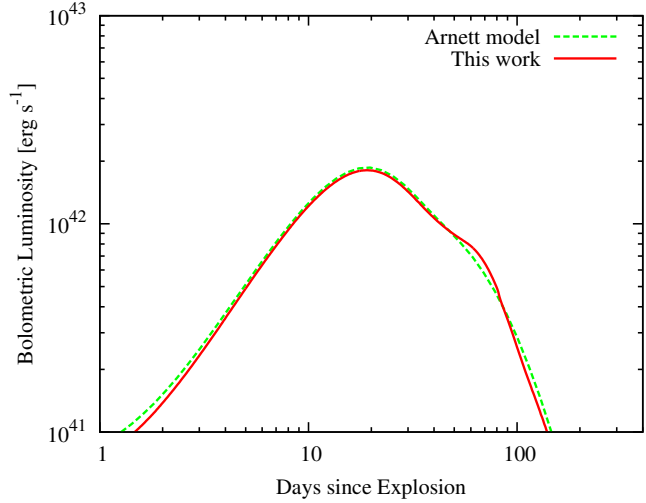}
\caption{
Bolometric SN light curves obtained by this work (solid red line) and the Arnett model (green-dashed line). 
Both lines are calculated for the same input physical parameters; ${\cal E}_{\rm K} = 1 \times 10^{51} \ \rm erg$, $M_{\rm ej} = 2 \ \rm M_\odot$, $M_{\rm {}^{56}Ni} = 0.1 \ \rm M_\odot$, and $K_{\rm T} = 0.1 \ \rm g^{-1} \ cm^{2}$. 
}
\label{fig:Arnett}
\end{figure}

The energy injection rate from the ${}^{56}$Ni decay is calculated as 
\begin{equation}
L_{\rm {}^{56}Ni} = M_{\rm {}^{56}Ni} \epsilon_{\rm {}^{56}Ni}\exp\left(-\frac{t}{t_{\rm {}^{56}Ni}}\right), 
\end{equation}
\begin{equation}
L_{\rm {}^{56}Co} = M_{\rm {}^{56}Ni} (\epsilon_{\rm {}^{56}Co}-\epsilon_{\rm {}^{56}Ni})\exp\left(-\frac{t}{t_{\rm {}^{56}Co}}\right),
\end{equation}
where $M_{\rm {}^{56}Ni}$ is the ${}^{56}$Ni mass,  $\epsilon_{\rm {}^{56}Ni} = 3.9 \times 10^{10} \ \rm erg \ s^{-1} \ g^{-1}$, 
$\epsilon_{\rm Co} = 6.8 \times 10^9 \ \rm erg \ s^{-1} \ g^{-1}$, $t_{\rm {}^{56}Ni} = 8.8 \ \rm days$, and $t_{\rm {}^{56}Co} = 111.3 \ \rm days$.
It is known that the energy deposition fraction from ${}^{56}$Ni decay can be well approximated by $f_{\rm dep, {}^{56}Ni} = 1- \exp(-\tau_{\rm eff})$  
with $\tau_{\rm eff} = (3-\delta) K_{\rm eff} M_{\rm ej}/4 \pi R_{\rm ej}{}^2$ and $K_{\rm eff} \approx 0.03 \ \rm cm^2 \ g^{-1}$. 
In this paper, we instead use Eq. (\ref{eq:f_dep}) to estimate the deposition fraction consistently with the wind dissipation, 
\begin{equation}
f_{\rm dep, {}^{56}Ni(Co)} = 
\frac{{\displaystyle \sum_{i}} f_{\rm dep}(\varepsilon_{{\rm {}^{56}Ni(Co)}, i}) {\cal P}_{{\rm {}^{56}Ni(Co)}, i} \varepsilon_{{\rm {}^{56}Ni(Co)}, i}}{{\displaystyle \sum_{i}} {\cal P}_{{\rm {}^{56}Ni(Co)}, i} \varepsilon_{{\rm {}^{56}Ni(Co)}, i}}. 
\end{equation}
where $\varepsilon_{{\rm {}^{56}Ni(Co)}, i}$ and ${\cal P}_{{\rm {}^{56}Ni(Co)}, i}$ are the mean energy of the decay product and the decay probability.
We consider 6 decay channels for ${}^{56}$Ni and 11 channels for ${}^{56}$Co listed in \cite{Nadyozhin_1994}. 
We assume that all the energy of positron emission goes to the thermal bath.  
As we show in Fig. \ref{fig:Arnett}, our model can broadly reproduce SN light curves obtained by using the conventional Arnett model, 
which confirms the validity of Eq. (\ref{eq:f_dep}).

%The key parameters, $E_{sn}$ and $M_{ej}$, can be determined from the peak time combined with the ejecta velocity is determined from the line features in the late-phase spectroscopy.  
%Also, the loaded nickel mass, $M_{{}^{56}{\rm Ni}}$ can be estimated from the peak flux. 
%Note that $E_{sn}$, or the initial kinetic energy of the ejecta,  can be derived by the above way only when the ejecta velocity is roughly constant, 
%i.e.,, when the energy injection via e.g., isotope decays is subdominant.  
%This corresponds to $M_{{}^{56}{\rm Ni}} \lesssim 1 M_\odot$, given that the decay energy is $\sim 60 \ \rm MeV$ per ${}^{56}$Ni, 
%which is a typical mass being synthesized during ordinary SN explosions.

%\bibliography{ref}

\begin{thebibliography}{109}
\expandafter\ifx\csname natexlab\endcsname\relax\def\natexlab#1{#1}\fi

\bibitem[{{Aartsen} {et~al.}(2015){Aartsen}, {Ackermann}, {Adams}, {Aguilar},
  {Ahlers}, {Ahrens}, {Altmann}, {Anderson}, {Arguelles}, {Arlen}, \&
  et~al.}]{Aartsen_et_al_2015}
{Aartsen}, M.~G., {Ackermann}, M., {Adams}, J., {et~al.} 2015, \apjl, 805, L5

\bibitem[{{Abe} {et~al.}(2011){Abe}, {Abe}, {Aihara}, {Fukuda}, {Hayato},
  {Huang}, {Ichikawa}, {Ikeda}, {Inoue}, {Ishino}, {Itow}, {Kajita}, {Kameda},
  {Kishimoto}, {Koga}, {Koshio}, {Lee}, {Minamino}, {Miura}, {Moriyama},
  {Nakahata}, {Nakamura}, {Nakaya}, {Nakayama}, {Nishijima}, {Nishimura},
  {Obayashi}, {Okumura}, {Sakuda}, {Sekiya}, {Shiozawa}, {Suzuki}, {Suzuki},
  {Takeda}, {Takeuchi}, {Tanaka}, {Tasaka}, {Tomura}, {Vagins}, {Wang}, \&
  {Yokoyama}}]{HK_2011}
{Abe}, K., {Abe}, T., {Aihara}, H., {et~al.} 2011, ArXiv e-prints

\bibitem[{{Abeysekara} {et~al.}(2012){Abeysekara}, {Aguilar}, {Aguilar},
  {Alfaro}, {Almaraz}, {{\'A}lvarez}, {{\'A}lvarez-Romero}, {{\'A}lvarez},
  {Arceo}, {Arteaga-Vel{\'a}zquez}, {Badillo}, {Barber}, {Baughman},
  {Bautista-Elivar}, {Belmont}, {Ben{\'{\i}}tez}, {BenZvi}, {Berley}, {Bernal},
  {Bonamente}, {Braun}, {Caballero-Lopez}, {Cabrera}, {Carrami{\~n}ana},
  {Carrasco}, {Castillo}, {Chambers}, {Conde}, {Condreay}, {Cotti}, {Cotzomi},
  {D'Olivo}, {de la Fuente}, {De Le{\'o}n}, {Delay}, {Delepine}, {DeYoung},
  {Diaz}, {Diaz-Cruz}, {Dingus}, {Duvernois}, {Edmunds}, {Ellsworth}, {Fick},
  {Fiorino}, {Flandes}, {Fraija}, {Galindo}, {Garc{\'{\i}}a-Luna},
  {Garc{\'{\i}}a-Torales}, {Garfias}, {Gonz{\'a}lez}, {Gonz{\'a}lez},
  {Goodman}, {Grabski}, {Gussert}, {Guzm{\'a}n-Ceron}, {Hampel-Arias},
  {Harris}, {Hays}, {Hernandez-Cervantes}, {H{\"u}ntemeyer}, {Imran},
  {Iriarte}, {Jimenez}, {Karn}, {Kelley-Hoskins}, {Kieda}, {Langarica}, {Lara},
  {Lauer}, {Lee}, {Linares}, {Linnemann}, {Longo}, {Luna-Garc{\'{\i}}a},
  {Mart{\'{\i}}nez}, {Mart{\'{\i}}nez}, {Mart{\'{\i}}nez}, {Mart{\'{\i}}nez},
  {Mart{\'{\i}}nez-Castro}, {Martos}, {Matthews}, {McEnery}, {Medina-Tanco},
  {Mendoza-Torres}, {Miranda-Romagnoli}, {Montaruli}, {Moreno}, {Mostafa},
  {Napsuciale}, {Nava}, {Nellen}, {Newbold}, {Noriega-Papaqui},
  {Oceguera-Becerra}, {Olmos Tapia}, {Orozco}, {P{\'e}rez},
  {P{\'e}rez-P{\'e}rez}, {Perkins}, {Pretz}, {Ramirez}, {Ram{\'{\i}}rez},
  {Rebello}, {Renter{\'{\i}}a}, {Reyes}, {Rosa-Gonz{\'a}lez}, {Rosado}, {Ryan},
  {Sacahui}, {Salazar}, {Salesa}, {Sandoval}, {Santos}, {Schneider}, {Shoup},
  {Silich}, {Sinnis}, {Smith}, {Sparks}, {Springer}, {Su{\'a}rez}, {Suarez},
  {Taboada}, {Tellez}, {Tenorio-Tagle}, {Tepe}, {Toale}, {Tollefson}, {Torres},
  {Ukwatta}, {Valdes-Galicia}, {Vanegas}, {Vasileiou}, {V{\'a}zquez},
  {V{\'a}zquez}, {Villase{\~n}or}, {Wall}, {Walters}, {Warner}, {Westerhoff},
  {Wisher}, {Wood}, {Yodh}, {Zaborov}, \& {Zepeda}}]{Abeysekara_et_al_2012}
{Abeysekara}, A.~U., {Aguilar}, J.~A., {Aguilar}, S., {et~al.} 2012,
  Astroparticle Physics, 35, 641

\bibitem[{{Accadia} {et~al.}(2011){Accadia}, {Acernese}, {Antonucci}, {Astone},
  {Ballardin}, {Barone}, {Barsuglia}, {Basti}, {Bauer}, {Bebronne}, {Beker},
  {Belletoile}, {Birindelli}, {Bitossi}, {Bizouard}, {Blom}, {Bondu},
  {Bonelli}, {Bonnand}, {Boschi}, {Bosi}, {Bouhou}, {Braccini}, {Bradaschia},
  {Branchesi}, {Briant}, {Brillet}, {Brisson}, {Budzy{\'n}ski}, {Bulik},
  {Bulten}, {Buskulic}, {Buy}, {Cagnoli}, {Calloni}, {Canuel}, {Carbognani},
  {Cavalier}, {Cavalieri}, {Cella}, {Cesarini}, {Chaibi}, {Chassande Mottin},
  {Chincarini}, {Cleva}, {Coccia}, {Cohadon}, {Colacino}, {Colas}, {Colla},
  {Colombini}, {Corsi}, {Coulon}, {Cuoco}, {D'Antonio}, {Dattilo}, {Davier},
  {Day}, {De Rosa}, {Debreczeni}, {Del Pozzo}, {del Prete}, {Di Fiore}, {Di
  Lieto}, {Emilio}, {Di Virgilio}, {Dietz}, {Drago}, {Fafone}, {Ferrante},
  {Fidecaro}, {Fiori}, {Flaminio}, {Forte}, {Fournier}, {Franc}, {Frasca},
  {Frasconi}, {Galimberti}, {Gammaitoni}, {Garufi}, {G{\'a}sp{\'a}r}, {Gemme},
  {Genin}, {Gennai}, {Giazotto}, {Gouaty}, {Granata}, {Greverie}, {Guidi},
  {Hayau}, {Heidmann}, {Heitmann}, {Hello}, {Huet}, {Jaranowski}, {Kowalska},
  {Kr{\'o}lak}, {Leroy}, {Letendre}, {Li}, {Liguori}, {Lorenzini}, {Loriette},
  {Losurdo}, {Majorana}, {Maksimovic}, {Man}, {Mantovani}, {Marchesoni},
  {Marion}, {Marque}, {Martelli}, {Masserot}, {Michel}, {Milano}, {Minenkov},
  {Mohan}, {Morgado}, {Morgia}, {Mosca}, {Moscatelli}, {Mours}, {Nocera},
  {Pagliaroli}, {Palladino}, {Palomba}, {Paoletti}, {Parisi}, {Pasqualetti},
  {Passaquieti}, {Passuello}, {Persichetti}, {Pichot}, {Piergiovanni},
  {Pietka}, {Pinard}, {Poggiani}, {Prato}, {Prodi}, {Punturo}, {Puppo},
  {Rabeling}, {R{\'a}cz}, {Rapagnani}, {Re}, {Regimbau}, {Ricci}, {Robinet},
  {Rocchi}, {Rolland}, {Romano}, {Rosi{\'n}ska}, {Ruggi}, {Sassolas},
  {Sentenac}, {Sperandio}, {Sturani}, {Swinkels}, {Tacca}, {Taffarello},
  {Toncelli}, {Tonelli}, {Torre}, {Tournefier}, {Travasso}, {Vajente}, {van den
  Brand}, {Van Den Broeck}, {van der Putten}, {Vasuth}, {Vavoulidis},
  {Vedovato}, {Verkindt}, {Vetrano}, {Vicer{\'e}}, {Vinet}, {Vitale}, {Vocca},
  {Ward}, {Was}, {Yvert}, \& {Zendri}}]{Accadia_et_al_2011}
{Accadia}, T., {Acernese}, F., {Antonucci}, F., {et~al.} 2011, Classical and
  Quantum Gravity, 28, 114002

\bibitem[{{Adams} {et~al.}(2013){Adams}, {Kochanek}, {Beacom}, {Vagins}, \&
  {Stanek}}]{Adams_et_al_2013}
{Adams}, S.~M., {Kochanek}, C.~S., {Beacom}, J.~F., {Vagins}, M.~R., \&
  {Stanek}, K.~Z. 2013, \apj, 778, 164

\bibitem[{{Akiyama} {et~al.}(2003){Akiyama}, {Wheeler}, {Meier}, \&
  {Lichtenstadt}}]{Akiyama_et_al_2003}
{Akiyama}, S., {Wheeler}, J.~C., {Meier}, D.~L., \& {Lichtenstadt}, I. 2003,
  \apj, 584, 954

\bibitem[{{Arnett}(1982)}]{Arnett_1982}
{Arnett}, W.~D. 1982, \apj, 253, 785

\bibitem[{{Atwood} {et~al.}(2009){Atwood}, {Abdo}, {Ackermann}, {Althouse},
  {Anderson}, {Axelsson}, {Baldini}, {Ballet}, {Band}, {Barbiellini}, \&
  et~al.}]{Atwood_et_al_2009}
{Atwood}, W.~B., {Abdo}, A.~A., {Ackermann}, M., {et~al.} 2009, \apj, 697, 1071

\bibitem[{{Balbus} \& {Hawley}(1998)}]{Balbus_Hawley_1998}
{Balbus}, S.~A., \& {Hawley}, J.~F. 1998, Reviews of Modern Physics, 70, 1

\bibitem[{{Bartos} {et~al.}(2013){Bartos}, {Brady}, \&
  {M{\'a}rka}}]{Bartos_et_al_2013}
{Bartos}, I., {Brady}, P., \& {M{\'a}rka}, S. 2013, Classical and Quantum
  Gravity, 30, 123001

\bibitem[{{Bibby} {et~al.}(2008){Bibby}, {Crowther}, {Furness}, \&
  {Clark}}]{Bibby_et_al_2008}
{Bibby}, J.~L., {Crowther}, P.~A., {Furness}, J.~P., \& {Clark}, J.~S. 2008,
  \mnras, 386, L23

\bibitem[{{Blackman} \& {Yi}(1998)}]{Blackman_Yi_1998}
{Blackman}, E.~G., \& {Yi}, I. 1998, \apjl, 498, L31

\bibitem[{{Bucciantini} {et~al.}(2007){Bucciantini}, {Quataert}, {Arons},
  {Metzger}, \& {Thompson}}]{Buccianitini_et_al_2007}
{Bucciantini}, N., {Quataert}, E., {Arons}, J., {Metzger}, B.~D., \&
  {Thompson}, T.~A. 2007, \mnras, 380, 1541

\bibitem[{{Bucciantini} {et~al.}(2008){Bucciantini}, {Quataert}, {Arons},
  {Metzger}, \& {Thompson}}]{Buccianitini_et_al_2008}
---. 2008, \mnras, 383, L25

\bibitem[{{Bucciantini} {et~al.}(2009){Bucciantini}, {Quataert}, {Metzger},
  {Thompson}, {Arons}, \& {Del Zanna}}]{Buccianitini_et_al_2009}
{Bucciantini}, N., {Quataert}, E., {Metzger}, B.~D., {et~al.} 2009, \mnras,
  396, 2038

\bibitem[{{Chatzopoulos} {et~al.}(2012){Chatzopoulos}, {Wheeler}, \&
  {Vinko}}]{Chatzopoulos_et_al_2012}
{Chatzopoulos}, E., {Wheeler}, J.~C., \& {Vinko}, J. 2012, \apj, 746, 121

\bibitem[{{Cutler}(2002)}]{Cutler_2002}
{Cutler}, C. 2002, \prd, 66, 084025

\bibitem[{{Cutler} \& {Jones}(2001)}]{Cutler_Jones_2001}
{Cutler}, C., \& {Jones}, D.~I. 2001, \prd, 63, 024002

\bibitem[{{Dall'Osso} {et~al.}(2009){Dall'Osso}, {Shore}, \&
  {Stella}}]{DallOsso_et_al_2009}
{Dall'Osso}, S., {Shore}, S.~N., \& {Stella}, L. 2009, \mnras, 398, 1869

\bibitem[{{Davies} {et~al.}(2009){Davies}, {Figer}, {Kudritzki}, {Trombley},
  {Kouveliotou}, \& {Wachter}}]{Davies_et_al_2009}
{Davies}, B., {Figer}, D.~F., {Kudritzki}, R.-P., {et~al.} 2009, \apj, 707, 844

\bibitem[{{Drout} {et~al.}(2011){Drout}, {Soderberg}, {Gal-Yam}, {Cenko},
  {Fox}, {Leonard}, {Sand}, {Moon}, {Arcavi}, \& {Green}}]{Drout_et_al_2011}
{Drout}, M.~R., {Soderberg}, A.~M., {Gal-Yam}, A., {et~al.} 2011, \apj, 741, 97

\bibitem[{{Duncan} \& {Thompson}(1992)}]{Duncan_Thompson_1992}
{Duncan}, R.~C., \& {Thompson}, C. 1992, \apjl, 392, L9

\bibitem[{{Eldridge} {et~al.}(2008){Eldridge}, {Izzard}, \&
  {Tout}}]{Eldridge_et_al_2008}
{Eldridge}, J.~J., {Izzard}, R.~G., \& {Tout}, C.~A. 2008, \mnras, 384, 1109

\bibitem[{{Fang} {et~al.}(2014){Fang}, {Kotera}, {Murase}, \&
  {Olinto}}]{Fang_et_al_2014}
{Fang}, K., {Kotera}, K., {Murase}, K., \& {Olinto}, A.~V. 2014, \prd, 90,
  103005

\bibitem[{{Faucher-Gigu{\`e}re} \& {Kaspi}(2006)}]{FaucherGigu_Kaspi_2006}
{Faucher-Gigu{\`e}re}, C.-A., \& {Kaspi}, V.~M. 2006, \apj, 643, 332

\bibitem[{{Ferrario} \& {Wickramasinghe}(2006)}]{Ferrario_Wickramasinghe_2006}
{Ferrario}, L., \& {Wickramasinghe}, D. 2006, \mnras, 367, 1323

\bibitem[{{Figer} {et~al.}(2005){Figer}, {Najarro}, {Geballe}, {Blum}, \&
  {Kudritzki}}]{Figer_et_al_2005}
{Figer}, D.~F., {Najarro}, F., {Geballe}, T.~R., {Blum}, R.~D., \& {Kudritzki},
  R.~P. 2005, \apjl, 622, L49

\bibitem[{{Fukuda}(1982)}]{Fukuda_1982}
{Fukuda}, I. 1982, \pasp, 94, 271

\bibitem[{{Gaensler} {et~al.}(2005){Gaensler}, {McClure-Griffiths}, {Oey},
  {Haverkorn}, {Dickey}, \& {Green}}]{Gaensler_et_al_2005}
{Gaensler}, B.~M., {McClure-Griffiths}, N.~M., {Oey}, M.~S., {et~al.} 2005,
  \apjl, 620, L95

\bibitem[{{Gal-Yam}(2012)}]{Gal-Yam_2012}
{Gal-Yam}, A. 2012, Science, 337, 927

\bibitem[{{Galama} {et~al.}(1998){Galama}, {Vreeswijk}, {van Paradijs},
  {Kouveliotou}, {Augusteijn}, {B{\"o}hnhardt}, {Brewer}, {Doublier},
  {Gonzalez}, {Leibundgut}, {Lidman}, {Hainaut}, {Patat}, {Heise}, {in't Zand},
  {Hurley}, {Groot}, {Strom}, {Mazzali}, {Iwamoto}, {Nomoto}, {Umeda},
  {Nakamura}, {Young}, {Suzuki}, {Shigeyama}, {Koshut}, {Kippen}, {Robinson},
  {de Wildt}, {Wijers}, {Tanvir}, {Greiner}, {Pian}, {Palazzi}, {Frontera},
  {Masetti}, {Nicastro}, {Feroci}, {Costa}, {Piro}, {Peterson}, {Tinney},
  {Boyle}, {Cannon}, {Stathakis}, {Sadler}, {Begam}, \&
  {Ianna}}]{Galama_et_al_1998}
{Galama}, T.~J., {Vreeswijk}, P.~M., {van Paradijs}, J., {et~al.} 1998, \nat,
  395, 670

\bibitem[{{Gehrels} {et~al.}(2004){Gehrels}, {Chincarini}, {Giommi}, {Mason},
  {Nousek}, {Wells}, {White}, {Barthelmy}, {Burrows}, {Cominsky}, {Hurley},
  {Marshall}, {M{\'e}sz{\'a}ros}, {Roming}, {Angelini}, {Barbier}, {Belloni},
  {Campana}, {Caraveo}, {Chester}, {Citterio}, {Cline}, {Cropper}, {Cummings},
  {Dean}, {Feigelson}, {Fenimore}, {Frail}, {Fruchter}, {Garmire}, {Gendreau},
  {Ghisellini}, {Greiner}, {Hill}, {Hunsberger}, {Krimm}, {Kulkarni}, {Kumar},
  {Lebrun}, {Lloyd-Ronning}, {Markwardt}, {Mattson}, {Mushotzky}, {Norris},
  {Osborne}, {Paczynski}, {Palmer}, {Park}, {Parsons}, {Paul}, {Rees},
  {Reynolds}, {Rhoads}, {Sasseen}, {Schaefer}, {Short}, {Smale}, {Smith},
  {Stella}, {Tagliaferri}, {Takahashi}, {Tashiro}, {Townsley}, {Tueller},
  {Turner}, {Vietri}, {Voges}, {Ward}, {Willingale}, {Zerbi}, \&
  {Zhang}}]{Gehrels_et_al_2004}
{Gehrels}, N., {Chincarini}, G., {Giommi}, P., {et~al.} 2004, \apj, 611, 1005

\bibitem[{{Greiner} {et~al.}(2008){Greiner}, {Bornemann}, {Clemens}, {Deuter},
  {Hasinger}, {Honsberg}, {Huber}, {Huber}, {Krauss}, {Kr{\"u}hler},
  {K{\"u}pc{\"u} Yolda{\c s}}, {Mayer-Hasselwander}, {Mican}, {Primak},
  {Schrey}, {Steiner}, {Szokoly}, {Th{\"o}ne}, {Yolda{\c s}}, {Klose}, {Laux},
  \& {Winkler}}]{Greiner_et_al_2008}
{Greiner}, J., {Bornemann}, W., {Clemens}, C., {et~al.} 2008, \pasp, 120, 405

\bibitem[{{Gruzinov}(2005)}]{Gruzinov_2005}
{Gruzinov}, A. 2005, Physical Review Letters, 94, 021101

\bibitem[{{Gualtieri} {et~al.}(2011){Gualtieri}, {Ciolfi}, \&
  {Ferrari}}]{Gualtieri_et_al_2011}
{Gualtieri}, L., {Ciolfi}, R., \& {Ferrari}, V. 2011, Classical and Quantum
  Gravity, 28, 114014

\bibitem[{{Guetta} \& {Della Valle}(2007)}]{Guetta_and_Della_2007}
{Guetta}, D., \& {Della Valle}, M. 2007, \apjl, 657, L73

\bibitem[{{Guetta} {et~al.}(2005){Guetta}, {Piran}, \&
  {Waxman}}]{Guetta_et_al_2005}
{Guetta}, D., {Piran}, T., \& {Waxman}, E. 2005, \apj, 619, 412

\bibitem[{{Harrison} {et~al.}(2013){Harrison}, {Craig}, {Christensen},
  {Hailey}, {Zhang}, {Boggs}, {Stern}, {Cook}, {Forster}, {Giommi},
  {Grefenstette}, {Kim}, {Kitaguchi}, {Koglin}, {Madsen}, {Mao}, {Miyasaka},
  {Mori}, {Perri}, {Pivovaroff}, {Puccetti}, {Rana}, {Westergaard}, {Willis},
  {Zoglauer}, {An}, {Bachetti}, {Barri{\`e}re}, {Bellm}, {Bhalerao},
  {Brejnholt}, {Fuerst}, {Liebe}, {Markwardt}, {Nynka}, {Vogel}, {Walton},
  {Wik}, {Alexander}, {Cominsky}, {Hornschemeier}, {Hornstrup}, {Kaspi},
  {Madejski}, {Matt}, {Molendi}, {Smith}, {Tomsick}, {Ajello}, {Ballantyne},
  {Balokovi{\'c}}, {Barret}, {Bauer}, {Blandford}, {Brandt}, {Brenneman},
  {Chiang}, {Chakrabarty}, {Chenevez}, {Comastri}, {Dufour}, {Elvis}, {Fabian},
  {Farrah}, {Fryer}, {Gotthelf}, {Grindlay}, {Helfand}, {Krivonos}, {Meier},
  {Miller}, {Natalucci}, {Ogle}, {Ofek}, {Ptak}, {Reynolds}, {Rigby},
  {Tagliaferri}, {Thorsett}, {Treister}, \& {Urry}}]{Harrison_et_al_2013}
{Harrison}, F.~A., {Craig}, W.~W., {Christensen}, F.~E., {et~al.} 2013, \apj,
  770, 103

\bibitem[{{Harry} \& {LIGO Scientific Collaboration}(2010)}]{Harry_et_al_2010}
{Harry}, G.~M., \& {LIGO Scientific Collaboration}. 2010, Classical and Quantum
  Gravity, 27, 084006

\bibitem[{{Heger} {et~al.}(2003){Heger}, {Fryer}, {Woosley}, {Langer}, \&
  {Hartmann}}]{Heter_et_al_2003}
{Heger}, A., {Fryer}, C.~L., {Woosley}, S.~E., {Langer}, N., \& {Hartmann},
  D.~H. 2003, \apj, 591, 288

\bibitem[{{Hodapp} {et~al.}(2004){Hodapp}, {Kaiser}, {Aussel}, {Burgett},
  {Chambers}, {Chun}, {Dombeck}, {Douglas}, {Hafner}, {Heasley}, {Hoblitt},
  {Hude}, {Isani}, {Jedicke}, {Jewitt}, {Laux}, {Luppino}, {Lupton}, {Maberry},
  {Magnier}, {Mannery}, {Monet}, {Morgan}, {Onaka}, {Price}, {Ryan},
  {Siegmund}, {Szapudi}, {Tonry}, {Wainscoat}, \&
  {Waterson}}]{Hodapp_et_al_2004}
{Hodapp}, K.~W., {Kaiser}, N., {Aussel}, H., {et~al.} 2004, Astronomische
  Nachrichten, 325, 636

\bibitem[{{Horiuchi} {et~al.}(2011){Horiuchi}, {Beacom}, {Kochanek}, {Prieto},
  {Stanek}, \& {Thompson}}]{Horiuchi_et_al_2011}
{Horiuchi}, S., {Beacom}, J.~F., {Kochanek}, C.~S., {et~al.} 2011, \apj, 738,
  154

\bibitem[{{Ikeda} {et~al.}(2007){Ikeda}, {Takeda}, {Fukuda}, {Vagins}, {Abe},
  {Iida}, {Ishihara}, {Kameda}, {Koshio}, {Minamino}, {Mitsuda}, {Miura},
  {Moriyama}, {Nakahata}, {Obayashi}, {Ogawa}, {Sekiya}, {Shiozawa}, {Suzuki},
  {Takeuchi}, {Ueshima}, {Watanabe}, {Yamada}, {Higuchi}, {Ishihara},
  {Ishitsuka}, {Kajita}, {Kaneyuki}, {Mitsuka}, {Nakayama}, {Nishino},
  {Okumura}, {Saji}, {Takenaga}, {Clark}, {Desai}, {Dufour}, {Kearns},
  {Likhoded}, {Litos}, {Raaf}, {Stone}, {Sulak}, {Wang}, {Goldhaber}, {Casper},
  {Cravens}, {Dunmore}, {Kropp}, {Liu}, {Mine}, {Regis}, {Smy}, {Sobel},
  {Ganezer}, {Hill}, {Keig}, {Jang}, {Kim}, {Lim}, {Scholberg}, {Tanimoto},
  {Walter}, {Wendell}, {Ellsworth}, {Tasaka}, {Guillian}, {Learned}, {Matsuno},
  {Messier}, {Hayato}, {Ichikawa}, {Ishida}, {Ishii}, {Iwashita}, {Kobayashi},
  {Nakadaira}, {Nakamura}, {Nitta}, {Oyama}, {Totsuka}, {Suzuki}, {Hasegawa},
  {Hiraide}, {Maesaka}, {Nakaya}, {Nishikawa}, {Sasaki}, {Yamamoto},
  {Yokoyama}, {Haines}, {Dazeley}, {Hatakeyama}, {Svoboda}, {Sullivan},
  {Turcan}, {Habig}, {Sato}, {Itow}, {Koike}, {Tanaka}, {Jung}, {Kato},
  {Kobayashi}, {Malek}, {McGrew}, {Sarrat}, {Terri}, {Yanagisawa}, {Tamura},
  {Idehara}, {Sakuda}, {Sugihara}, {Kuno}, {Yoshida}, {Kim}, {Yang}, {Yoo},
  {Ishizuka}, {Okazawa}, {Choi}, {Seo}, {Gando}, {Hasegawa}, {Inoue}, {Furuse},
  {Ishii}, {Nishijima}, {Ishino}, {Watanabe}, {Koshiba}, {Chen}, {Deng}, {Liu},
  {Kielczewska}, {Zalipska}, {Berns}, {Gran}, {Shiraishi}, {Stachyra},
  {Thrane}, {Washburn}, {Wilkes}, \& {Super-KAMIOKANDE
  Collaboration}}]{Ikeda_et_al_2007}
{Ikeda}, M., {Takeda}, A., {Fukuda}, Y., {et~al.} 2007, \apj, 669, 519

\bibitem[{{Inserra} {et~al.}(2013){Inserra}, {Smartt}, {Jerkstrand}, {Valenti},
  {Fraser}, {Wright}, {Smith}, {Chen}, {Kotak}, {Pastorello}, {Nicholl},
  {Bresolin}, {Kudritzki}, {Benetti}, {Botticella}, {Burgett}, {Chambers},
  {Ergon}, {Flewelling}, {Fynbo}, {Geier}, {Hodapp}, {Howell}, {Huber},
  {Kaiser}, {Leloudas}, {Magill}, {Magnier}, {McCrum}, {Metcalfe}, {Price},
  {Rest}, {Sollerman}, {Sweeney}, {Taddia}, {Taubenberger}, {Tonry},
  {Wainscoat}, {Waters}, \& {Young}}]{Inserra_et_al_2013}
{Inserra}, C., {Smartt}, S.~J., {Jerkstrand}, A., {et~al.} 2013, \apj, 770, 128

\bibitem[{{Kasen} \& {Bildsten}(2010)}]{Kasen_Bildsten_2010}
{Kasen}, D., \& {Bildsten}, L. 2010, \apj, 717, 245

\bibitem[{{Kashiyama} {et~al.}(2013){Kashiyama}, {Murase}, {Horiuchi}, {Gao},
  \& {M{\'e}sz{\'a}ros}}]{Kashiyama_et_al_2013}
{Kashiyama}, K., {Murase}, K., {Horiuchi}, S., {Gao}, S., \&
  {M{\'e}sz{\'a}ros}, P. 2013, \apjl, 769, L6

\bibitem[{{Katz} {et~al.}(2011){Katz}, {Sapir}, \& {Waxman}}]{Katz_et_al_2011}
{Katz}, B., {Sapir}, N., \& {Waxman}, E. 2011, ArXiv e-prints

\bibitem[{{Keane} \& {Kramer}(2008)}]{Keane_Kramer_2008}
{Keane}, E.~F., \& {Kramer}, M. 2008, \mnras, 391, 2009

\bibitem[{{Kistler} {et~al.}(2013){Kistler}, {Haxton}, \&
  {Y{\"u}ksel}}]{Kistler_et_al_2013}
{Kistler}, M.~D., {Haxton}, W.~C., \& {Y{\"u}ksel}, H. 2013, \apj, 778, 81

\bibitem[{{Kiuchi} \& {Yoshida}(2008)}]{Kiuchi_Yoshida_2008}
{Kiuchi}, K., \& {Yoshida}, S. 2008, \prd, 78, 044045

\bibitem[{{Kokkotas}(2008)}]{Kokkotas_2008}
{Kokkotas}, K.~D. 2008, in Reviews in Modern Astronomy, Vol.~20, Reviews in
  Modern Astronomy, ed. S.~{R{\"o}ser}, 140

\bibitem[{{Kotera} {et~al.}(2013){Kotera}, {Phinney}, \&
  {Olinto}}]{Kotera_et_al_2013}
{Kotera}, K., {Phinney}, E.~S., \& {Olinto}, A.~V. 2013, \mnras, 432, 3228

\bibitem[{{Lattimer} \& {Prakash}(2001)}]{Lattimer_Prakash_2001}
{Lattimer}, J.~M., \& {Prakash}, M. 2001, \apj, 550, 426

\bibitem[{{Law} {et~al.}(2009){Law}, {Kulkarni}, {Dekany}, {Ofek}, {Quimby},
  {Nugent}, {Surace}, {Grillmair}, {Bloom}, {Kasliwal}, {Bildsten}, {Brown},
  {Cenko}, {Ciardi}, {Croner}, {Djorgovski}, {van Eyken}, {Filippenko}, {Fox},
  {Gal-Yam}, {Hale}, {Hamam}, {Helou}, {Henning}, {Howell}, {Jacobsen},
  {Laher}, {Mattingly}, {McKenna}, {Pickles}, {Poznanski}, {Rahmer}, {Rau},
  {Rosing}, {Shara}, {Smith}, {Starr}, {Sullivan}, {Velur}, {Walters}, \&
  {Zolkower}}]{Law_et_al_2009}
{Law}, N.~M., {Kulkarni}, S.~R., {Dekany}, R.~G., {et~al.} 2009, \pasp, 121,
  1395

\bibitem[{{Lemoine} {et~al.}(2015){Lemoine}, {Kotera}, \&
  {P{\'e}tri}}]{Lemoine_et_al_2015}
{Lemoine}, M., {Kotera}, K., \& {P{\'e}tri}, J. 2015, JCAP, 7, 16

\bibitem[{{LIGO Scientific Collaboration} {et~al.}(2013){LIGO Scientific
  Collaboration}, {Virgo Collaboration}, {Aasi}, {Abadie}, {Abbott}, {Abbott},
  {Abbott}, {Abernathy}, {Accadia}, {Acernese}, \& et~al.}]{LIGO_Virgo_2013}
{LIGO Scientific Collaboration}, {Virgo Collaboration}, {Aasi}, J., {et~al.}
  2013, ArXiv e-prints

\bibitem[{{Lyman} {et~al.}(2014){Lyman}, {Bersier}, {James}, {Mazzali},
  {Eldridge}, {Fraser}, \& {Pian}}]{Lyman_et_al_2014}
{Lyman}, J., {Bersier}, D., {James}, P., {et~al.} 2014, ArXiv e-prints

\bibitem[{{Maeda}(2006)}]{Maeda_2006}
{Maeda}, K. 2006, \apj, 644, 385

\bibitem[{{Maeda} {et~al.}(2007){Maeda}, {Tanaka}, {Nomoto}, {Tominaga},
  {Kawabata}, {Mazzali}, {Umeda}, {Suzuki}, \& {Hattori}}]{Maeda_et_al_2007}
{Maeda}, K., {Tanaka}, M., {Nomoto}, K., {et~al.} 2007, \apj, 666, 1069

\bibitem[{{Makishima} {et~al.}(2014){Makishima}, {Enoto}, {Hiraga}, {Nakano},
  {Nakazawa}, {Sakurai}, {Sasano}, \& {Murakami}}]{Makishima_et_al_2014}
{Makishima}, K., {Enoto}, T., {Hiraga}, J.~S., {et~al.} 2014, Physical Review
  Letters, 112, 171102

\bibitem[{{Makishima} {et~al.}(2015){Makishima}, {Enoto}, {Murakami}, {Furuta},
  {Nakano}, {Sasano}, \& {Nakazawa}}]{Makishima_et_al_2015}
{Makishima}, K., {Enoto}, T., {Murakami}, H., {et~al.} 2015, \pasj

\bibitem[{{Mazzali} {et~al.}(2006){Mazzali}, {Deng}, {Nomoto}, {Sauer}, {Pian},
  {Tominaga}, {Tanaka}, {Maeda}, \& {Filippenko}}]{Mazzali_et_al_2006}
{Mazzali}, P.~A., {Deng}, J., {Nomoto}, K., {et~al.} 2006, \nat, 442, 1018

\bibitem[{{Mestel} \& {Takhar}(1972)}]{Mestel_Takhar_1972}
{Mestel}, L., \& {Takhar}, H.~S. 1972, \mnras, 156, 419

\bibitem[{{Metzger} {et~al.}(2015){Metzger}, {Margalit}, {Kasen}, \&
  {Quataert}}]{Metzger_et_al_2015}
{Metzger}, B.~D., {Margalit}, B., {Kasen}, D., \& {Quataert}, E. 2015, \mnras,
  454, 3311

\bibitem[{{Metzger} {et~al.}(2007){Metzger}, {Thompson}, \&
  {Quataert}}]{Metzger_et_al_2007}
{Metzger}, B.~D., {Thompson}, T.~A., \& {Quataert}, E. 2007, \apj, 659, 561

\bibitem[{{Metzger} {et~al.}(2014){Metzger}, {Vurm}, {Hasco{\"e}t}, \&
  {Beloborodov}}]{Metzger_et_al_2013}
{Metzger}, B.~D., {Vurm}, I., {Hasco{\"e}t}, R., \& {Beloborodov}, A.~M. 2014,
  \mnras, 437, 703

\bibitem[{{Meynet} {et~al.}(2011){Meynet}, {Eggenberger}, \&
  {Maeder}}]{Meynet_et_al_2011}
{Meynet}, G., {Eggenberger}, P., \& {Maeder}, A. 2011, \aap, 525, L11

\bibitem[{{M{\"o}sta} {et~al.}(2015){M{\"o}sta}, {Ott}, {Radice}, {Roberts},
  {Schnetter}, \& {Haas}}]{Mosta_et_al_2015}
M{\"o}sta P., Ott C.~D., Radice D., Roberts L.~F., Schnetter E., Haas R., 2015, Natur, 528, 376

\bibitem[{{Muno} {et~al.}(2006){Muno}, {Clark}, {Crowther}, {Dougherty}, {de
  Grijs}, {Law}, {McMillan}, {Morris}, {Negueruela}, {Pooley}, {Portegies
  Zwart}, \& {Yusef-Zadeh}}]{Muno_et_al_2006}
{Muno}, M.~P., {Clark}, J.~S., {Crowther}, P.~A., {et~al.} 2006, \apjl, 636,
  L41

\bibitem[{{Murase} {et~al.}(2014){Murase}, {Dasgupta}, \&
  {Thompson}}]{Murase_et_al_2014}
{Murase}, K., {Dasgupta}, B., \& {Thompson}, T.~A. 2014, \prd, 89, 043012

\bibitem[{{Murase} {et~al.}(2015){Murase}, {Kashiyama}, {Kiuchi}, \&
  {Bartos}}]{Murase_et_al_2015}
{Murase}, K., {Kashiyama}, K., {Kiuchi}, K., \& {Bartos}, I. 2015, \apj, 805,
  82

\bibitem[{{Murase} {et~al.}(2009){Murase}, {M{\'e}sz{\'a}ros}, \&
  {Zhang}}]{Murase_et_al_2009}
{Murase}, K., {M{\'e}sz{\'a}ros}, P., \& {Zhang}, B. 2009, \prd, 79, 103001

\bibitem[{{Murase} {et~al.}(2011){Murase}, {Thompson}, {Lacki}, \&
  {Beacom}}]{Murase_et_al_2011}
{Murase}, K., {Thompson}, T.~A., {Lacki}, B.~C., \& {Beacom}, J.~F. 2011, \prd,
  84, 043003

\bibitem[{{Nadyozhin}(1994)}]{Nadyozhin_1994}
{Nadyozhin}, D.~K. 1994, \apjs, 92, 527

\bibitem[{{Nakauchi} {et~al.}(2015){Nakauchi}, {Kashiyama}, {Nagakura}, {Suwa},
  \& {Nakamura}}]{Nakauchi_et_al_2015}
{Nakauchi}, D., {Kashiyama}, K., {Nagakura}, H., {Suwa}, Y., \& {Nakamura}, T.
  2015, \apj, 805, 164

\bibitem[{{Nicholl} {et~al.}(2013){Nicholl}, {Smartt}, {Jerkstrand}, {Inserra},
  {McCrum}, {Kotak}, {Fraser}, {Wright}, {Chen}, {Smith}, {Young}, {Sim},
  {Valenti}, {Howell}, {Bresolin}, {Kudritzki}, {Tonry}, {Huber}, {Rest},
  {Pastorello}, {Tomasella}, {Cappellaro}, {Benetti}, {Mattila}, {Kankare},
  {Kangas}, {Leloudas}, {Sollerman}, {Taddia}, {Berger}, {Chornock}, {Narayan},
  {Stubbs}, {Foley}, {Lunnan}, {Soderberg}, {Sanders}, {Milisavljevic},
  {Margutti}, {Kirshner}, {Elias-Rosa}, {Morales-Garoffolo}, {Taubenberger},
  {Botticella}, {Gezari}, {Urata}, {Rodney}, {Riess}, {Scolnic}, {Wood-Vasey},
  {Burgett}, {Chambers}, {Flewelling}, {Magnier}, {Kaiser}, {Metcalfe},
  {Morgan}, {Price}, {Sweeney}, \& {Waters}}]{Nicholl_et_al_2013}
{Nicholl}, M., {Smartt}, S.~J., {Jerkstrand}, A., {et~al.} 2013, \nat, 502, 346

\bibitem[{{Olausen} \& {Kaspi}(2014)}]{Olausen_Kaspi_2014}
{Olausen}, S.~A., \& {Kaspi}, V.~M. 2014, \apjs, 212, 6

\bibitem[{{Ostriker} \& {Gunn}(1969)}]{Ostriker_Gunn_1969}
{Ostriker}, J.~P., \& {Gunn}, J.~E. 1969, \apj, 157, 1395

\bibitem[{{Owen} \& {Lindblom}(2002)}]{Owen_Lindblom_2002}
{Owen}, B.~J., \& {Lindblom}, L. 2002, Classical and Quantum Gravity, 19, 1247

\bibitem[{{Pastorello} {et~al.}(2010){Pastorello}, {Smartt}, {Botticella},
  {Maguire}, {Fraser}, {Smith}, {Kotak}, {Magill}, {Valenti}, {Young},
  {Gezari}, {Bresolin}, {Kudritzki}, {Howell}, {Rest}, {Metcalfe}, {Mattila},
  {Kankare}, {Huang}, {Urata}, {Burgett}, {Chambers}, {Dombeck}, {Flewelling},
  {Grav}, {Heasley}, {Hodapp}, {Kaiser}, {Luppino}, {Lupton}, {Magnier},
  {Monet}, {Morgan}, {Onaka}, {Price}, {Rhoads}, {Siegmund}, {Stubbs},
  {Sweeney}, {Tonry}, {Wainscoat}, {Waterson}, {Waters}, \&
  {Wynn-Williams}}]{Pastorello_et_al_2010}
{Pastorello}, A., {Smartt}, S.~J., {Botticella}, M.~T., {et~al.} 2010, \apjl,
  724, L16

\bibitem[{{Perley} {et~al.}(2011){Perley}, {Chandler}, {Butler}, \&
  {Wrobel}}]{Perley_et_al_2011}
{Perley}, R.~A., {Chandler}, C.~J., {Butler}, B.~J., \& {Wrobel}, J.~M. 2011,
  \apjl, 739, L1

\bibitem[{{Perna} {et~al.}(2008){Perna}, {Soria}, {Pooley}, \&
  {Stella}}]{Perna_et_al_2008}
{Perna}, R., {Soria}, R., {Pooley}, D., \& {Stella}, L. 2008, \mnras, 384, 1638

\bibitem[{{Piro} \& {Thrane}(2012)}]{Piro_Thrane_2012}
{Piro}, A.~L., \& {Thrane}, E. 2012, \apj, 761, 63

\bibitem[{{Popov} {et~al.}(2010){Popov}, {Pons}, {Miralles}, {Boldin}, \&
  {Posselt}}]{Popov_et_al_2010}
{Popov}, S.~B., {Pons}, J.~A., {Miralles}, J.~A., {Boldin}, P.~A., \&
  {Posselt}, B. 2010, \mnras, 401, 2675

\bibitem[{{Quimby} {et~al.}(2013){Quimby}, {Yuan}, {Akerlof}, \&
  {Wheeler}}]{Quimby_et_al_2013}
{Quimby}, R.~M., {Yuan}, F., {Akerlof}, C., \& {Wheeler}, J.~C. 2013, \mnras,
  431, 912

\bibitem[{{Quimby} {et~al.}(2011){Quimby}, {Kulkarni}, {Kasliwal}, {Gal-Yam},
  {Arcavi}, {Sullivan}, {Nugent}, {Thomas}, {Howell}, {Nakar}, {Bildsten},
  {Theissen}, {Law}, {Dekany}, {Rahmer}, {Hale}, {Smith}, {Ofek}, {Zolkower},
  {Velur}, {Walters}, {Henning}, {Bui}, {McKenna}, {Poznanski}, {Cenko}, \&
  {Levitan}}]{Quimby_et_al_2011}
{Quimby}, R.~M., {Kulkarni}, S.~R., {Kasliwal}, M.~M., {et~al.} 2011, \nat,
  474, 487

\bibitem[{{Shappee} {et~al.}(2014){Shappee}, {Prieto}, {Grupe}, {Kochanek},
  {Stanek}, {De Rosa}, {Mathur}, {Zu}, {Peterson}, {Pogge}, {Komossa}, {Im},
  {Jencson}, {Holoien}, {Basu}, {Beacom}, {Szczygie{\l}}, {Brimacombe},
  {Adams}, {Campillay}, {Choi}, {Contreras}, {Dietrich}, {Dubberley},
  {Elphick}, {Foale}, {Giustini}, {Gonzalez}, {Hawkins}, {Howell}, {Hsiao},
  {Koss}, {Leighly}, {Morrell}, {Mudd}, {Mullins}, {Nugent}, {Parrent},
  {Phillips}, {Pojmanski}, {Rosing}, {Ross}, {Sand}, {Terndrup}, {Valenti},
  {Walker}, \& {Yoon}}]{Shappee_et_al_2014}
{Shappee}, B.~J., {Prieto}, J.~L., {Grupe}, D., {et~al.} 2014, \apj, 788, 48

\bibitem[{{Smith} {et~al.}(2011){Smith}, {Li}, {Filippenko}, \&
  {Chornock}}]{Smith_et_al_2011}
{Smith}, N., {Li}, W., {Filippenko}, A.~V., \& {Chornock}, R. 2011, \mnras,
  412, 1522

\bibitem[{{Soderberg} {et~al.}(2006){Soderberg}, {Kulkarni}, {Nakar}, {Berger},
  {Cameron}, {Fox}, {Frail}, {Gal-Yam}, {Sari}, {Cenko}, {Kasliwal},
  {Chevalier}, {Piran}, {Price}, {Schmidt}, {Pooley}, {Moon}, {Penprase},
  {Ofek}, {Rau}, {Gehrels}, {Nousek}, {Burrows}, {Persson}, \&
  {McCarthy}}]{Soderberg_et_al_2006}
{Soderberg}, A.~M., {Kulkarni}, S.~R., {Nakar}, E., {et~al.} 2006, \nat, 442,
  1014

\bibitem[{{Sollerman} {et~al.}(2002){Sollerman}, {Holland}, {Challis},
  {Fransson}, {Garnavich}, {Kirshner}, {Kozma}, {Leibundgut}, {Lundqvist},
  {Patat}, {Filippenko}, {Panagia}, \& {Wheeler}}]{Sollerman_et_al_2002}
{Sollerman}, J., {Holland}, S.~T., {Challis}, P., {et~al.} 2002, \aap, 386, 944

\bibitem[{{Somiya}(2012)}]{Somiya_2012}
{Somiya}, K. 2012, Classical and Quantum Gravity, 29, 124007

\bibitem[{{Spitkovsky}(2006)}]{Spitkovsky_2006}
{Spitkovsky}, A. 2006, \apjl, 648, L51

\bibitem[{{Stella} {et~al.}(2005){Stella}, {Dall'Osso}, {Israel}, \&
  {Vecchio}}]{Stella_et_al_2005}
{Stella}, L., {Dall'Osso}, S., {Israel}, G.~L., \& {Vecchio}, A. 2005, \apjl,
  634, L165

\bibitem[{{Tanaka} \& {Takahara}(2010)}]{Tanaka_Takahara_2010}
{Tanaka}, S.~J., \& {Takahara}, F. 2010, \apj, 715, 1248

\bibitem[{{Tchekhovskoy} {et~al.}(2013){Tchekhovskoy}, {Spitkovsky}, \&
  {Li}}]{Tchekohovskoy_et_al_2013}
{Tchekhovskoy}, A., {Spitkovsky}, A., \& {Li}, J.~G. 2013, \mnras, 435, L1

\bibitem[{{Thompson}(1994)}]{Thompson_1994}
{Thompson}, C. 1994, \mnras, 270, 480

\bibitem[{{Thompson} \& {Duncan}(1993)}]{Thompson_Duncan_1993}
{Thompson}, C., \& {Duncan}, R.~C. 1993, \apj, 408, 194

\bibitem[{{Thompson} {et~al.}(2004){Thompson}, {Chang}, \&
  {Quataert}}]{Thompson_et_al_2004}
{Thompson}, T.~A., {Chang}, P., \& {Quataert}, E. 2004, \apj, 611, 380

\bibitem[{{Thompson} {et~al.}(2005){Thompson}, {Quataert}, \&
  {Burrows}}]{Thompson_et_al_2005}
{Thompson}, T.~A., {Quataert}, E., \& {Burrows}, A. 2005, \apj, 620, 861

\bibitem[{{Thrane} {et~al.}(2011){Thrane}, {Kandhasamy}, {Ott}, {Anderson},
  {Christensen}, {Coughlin}, {Dorsher}, {Giampanis}, {Mandic}, {Mytidis},
  {Prestegard}, {Raffai}, \& {Whiting}}]{Thrane_et_al_2011}
{Thrane}, E., {Kandhasamy}, S., {Ott}, C.~D., {et~al.} 2011, \prd, 83, 083004

\bibitem[{{Toma} {et~al.}(2007){Toma}, {Ioka}, {Sakamoto}, \&
  {Nakamura}}]{Toma_et_al_2007}
{Toma}, K., {Ioka}, K., {Sakamoto}, T., \& {Nakamura}, T. 2007, \apj, 659, 1420

\bibitem[{{Usov}(1992)}]{Usov_1992}
{Usov}, V.~V. 1992, \nat, 357, 472

\bibitem[{{Vink} \& {Kuiper}(2006)}]{Vink_Kuiper_2006}
{Vink}, J., \& {Kuiper}, L. 2006, \mnras, 370, L14

\bibitem[{{Wanderman} \& {Piran}(2010)}]{Wanderman_Piran_2010}
{Wanderman}, D., \& {Piran}, T. 2010, \mnras, 406, 1944

\bibitem[{{Wang} {et~al.}(2015){Wang}, {Wang}, {Dai}, \&
  {Wu}}]{Wang_et_al_2015}
{Wang}, S.~Q., {Wang}, L.~J., {Dai}, Z.~G., \& {Wu}, X.~F. 2015, \apj, 807, 147

\bibitem[{{Wheeler} {et~al.}(2015){Wheeler}, {Johnson}, \&
  {Clocchiatti}}]{Wheeler_et_al_2015}
{Wheeler}, J.~C., {Johnson}, V., \& {Clocchiatti}, A. 2015, \mnras, 450, 1295

\bibitem[{{Wheeler} {et~al.}(2000){Wheeler}, {Yi}, {H{\"o}flich}, \&
  {Wang}}]{Wheeler_et_al_2000}
{Wheeler}, J.~C., {Yi}, I., {H{\"o}flich}, P., \& {Wang}, L. 2000, \apj, 537,
  810

\bibitem[{{Woosley}(2010)}]{Woosley_2010}
{Woosley}, S.~E. 2010, \apjl, 719, L204

\bibitem[{{Zhang} \& {M{\'e}sz{\'a}ros}(2001)}]{Zhang_Meszaros_2001}
{Zhang}, B., \& {M{\'e}sz{\'a}ros}, P. 2001, \apjl, 552, L35

\end{thebibliography}

\end{document}